\documentclass{article}
\usepackage {amsmath}
\usepackage {amssymb}
\usepackage {mathabx}
\usepackage {amsthm}
\usepackage {xcolor}
\usepackage {tikz}
\usepackage {tikz-cd}
\usetikzlibrary {graphs}
\usepackage [framemethod=tikz] {mdframed}
\usepackage {algorithm}
\usepackage {algorithmic}
\usepackage {float}
\usepackage {fancybox}
\usepackage {CJKutf8}
\usepackage {graphicx}
\usepackage {anyfontsize}
\usepackage[title]{appendix}

\makeatletter
\newenvironment{breakablealgorithm}
  {
   \begin{center}
     \refstepcounter{algorithm}
     \hrule height.8pt depth0pt \kern2pt
     \renewcommand{\caption}[2][\relax]{
       {\raggedright\textbf{\ALG@name~\thealgorithm} ##2\par}
       \ifx\relax##1\relax
         \addcontentsline{loa}{algorithm}{\protect\numberline{\thealgorithm}##2}
       \else
         \addcontentsline{loa}{algorithm}{\protect\numberline{\thealgorithm}##1}
       \fi
       \kern2pt\hrule\kern2pt
     }
  }{
     \kern2pt\hrule\relax
   \end{center}
  }
\makeatother

\begin{document}
\begin{CJK}{UTF8}{gbsn}

\title{A proof of P$\neq$NP (New symmetric encryption algorithm against any linear attacks and differential attacks)}
\author{Gao Ming\thanks{20070602094@alu.cqu.edu.cn}}
\date{March 5, 2022}

\begin{titlepage}

\maketitle

\vspace{7mm}
\begin{abstract}
P  vs  NP  problem  is  the   most  important  unresolved  problem  in  the  field  of computational complexity. Its impact has penetrated into all aspects of algorithm design,  especially  in  the  field  of  cryptography.  The  security  of  cryptographic algorithms  based  on  short  keys  depends  on  whether  P  is  equal  to  NP. In fact, the security requirements for cryptographic keys are much stricter than those for P$\neq$NP, the security of the key must ensure not only a sufficiently high computational complexity to crack it, but also consider the security of each bit of the key, while fully avoiding the effectiveness of various attack methods.\\
\indent
In this paper, we innovatively propose a new encoding mechanism and develop a novel block symmetric encryption algorithm, which be named Eagle,  whose encryption and decryption can be completed in linear time. The key consists of 12 variables, for the attacker, in the case when only the plaintext-ciphertext correspondence is known, the problem of cracking the key is equivalent to solving a system of equations about 12 unknown variables $(w_{1}, w_{2}, ..., w_{11}, w_{12})$ together with some implicit variables which are generated by the encryption process. \\
\indent
In this paper, we will prove that there is no universal algorithm without brute-force searching at least one variable during the process of verifying five variables $(w_{1}$, $w_{2}$, $w_{3}$, $w_{4}$, $w_{9})$ whose computational complexity is not greater than solving the 12 variables. After performing elimination the implicit variables, the problem of verifying the five variables will be equivalently transformed into a series of general quintic equations about one variable $w_{11} \oplus w_{12}$ with coefficient given by other six variables $(w_{5}$, $w_{6}$, $w_{7}$, $w_{8}$, $w_{10}$, $w_{11})$, according to the conclusion that ``there is no algebraic solution to quintic equations", we can't establish a system of algebraic equations only containing the six variables $(w_{5}$, $w_{6}$, $w_{7}$, $w_{8}$, $w_{10}$, $w_{11})$ without $w_{11} \oplus w_{12}$. Without solving the six variables, the problem of verifying the five variables $(w_{1}$, $w_{2}$, $w_{3}$, $w_{4}$, $w_{9})$ is equivalent to solve the slope of a line given only the x-coordinates of several points on it, this implies that there is no universal algorithm without brute-force searching at least one variable to verify the five variables $(w_{1}$, $w_{2}$, $w_{3}$, $w_{4}$, $w_{9})$. Since the computational complexity of brute-force searching any single variable is exponential, thus the complexity of cracking the key is exponential. Thus decryption is a one-way function, and according to ``the existence of one-way function means P$\neq$NP", solving the unsolved problem of P vs NP. \\
\indent
In addition, we further constructed another block symmetric encryption algorithm, which be named $Eagle^{*}$. The key of $Eagle^{*}$ is composed of three independent variables. Given any plaintext ciphertext correspondences, the problem of verifying any two variables in the key is equivalent to solving a system of quintic equations. Based on the conclusion that there is no algebraic solution to quintic equations, we assert that there is no fast algorithm to verify any two variables, and thus the computational complexity of cracking the key is equal to the computational complexity of completely exhausting the three variables. In addition, we conducted an in-depth analysis that without cracking the key, it is not possible to directly decrypt new ciphertext by simply collecting a large number of ciphertext plaintext correspondences. Due to the fact that ciphertext consists of four parts, three parts are completely random, while the other small part is calculated using plaintexts, keys, and random numbers, and their distribution is also completely random, thus $Eagle^{*}$ is resistant to all forms of linear and differential attacks. \\
\end{abstract}

\end{titlepage}

\newpage

\section{Introduction}
\indent
Cryptography   is   one   of   the   most   important   applications   in   the   field   of communication  and  computer  science.  In  recent  years,  with  the  application  of commerce,  enterprises,  banks  and  other  departments,  cryptography  has  been developed rapidly. Especially after Shannon put forward the mathematical analysis of security in ``Communication theory of secrecy systems"\cite{REF1}, various design tools for cipher algorithms and corresponding attack tools have been developed one after  another.  \\
\indent
The P vs NP problem was first proposed by Cook\cite{REF10} and has become one of the most important unsolved problems in the field of computer science. At the same time, the P vs NP problem has also become the most important theoretical foundation in the field of cryptography, determining the research direction of cryptography. P$\neq$NP is the most basic assumption in cryptography, but this is only a necessary condition. In fact, the security requirements of a key are much stricter than P$\neq$NP. This is because even if P$\neq$NP, there exists an  encryption algorithm, the computational complexity of cracking the key is exponential to the length of the key, which does not guarantee that each bit of the key is secure. To ensure that each bit of the key is secure, it is necessary to ensure that exhaustive search is only effective for cracking the key, which is much stricter than the exponential computational complexity to cracking the key.\\
\indent
Among various attack methods, linear attack and differential attack are the two most core attack methods, which have become important analytical tools in the field of cryptography, driving the improvement of encryption algorithm.\\
\indent
Linear attack was first proposed by M. Matsui\cite{REF2}, this is an attack method that is currently  applicable  to  almost   all   block  encryption  algorithms.   Kaliski  BS\cite{REF3} proposed  a  multi-linear  attack  based  on  the  linear  attack,  but  the  multi-linear attack has many limitations. And the Biryukov A\cite{REF4} and Chao, J.Y\cite{REF5} and others further improved the framework of multi-linear attacks, thus making linear attacks a larger application.\\
\indent
The differential attack method was first proposed by Eli  Biham\cite{REF6}.  Biham  E\cite{REF7} extended it to a more powerful attack method. Tsunoo\cite{REF8} further constructed multiple attack methods. These attack methods  have extremely high skill in the attack process, which is worthy in-depth study.\\
\indent
In Chapter 3, we propose a novel encoding mechanism which we named Eagle. The distinguishing feature of Eagle encoding algorithm from any other encoding algorithm is that for any given group of input and output texts, the key parameters are unsolvable. \\
\indent
In Chapter 4, we further delve into the mathematical laws behind the Eagle encoding algorithm and discover an operation that satisfies the allocation law for XOR operations. This operation is very similar to the allocation law of multiplication operations, but does not satisfy the commutative and associative laws, we refer to it as a ``right multiplication operation on binary". Then we construct a binary operation of two-dimensional vectors which serves as a key ingredient for the subsequent design of encryption algorithms. \\
\\
\indent
In Chapter 5, we constructed a block symmetric encryption algorithm based on the Eagle encoding algorithm which we named it Eagle. In this encryption algorithm, there are six completely independent sets of random keys, each containing two random variables. For the plaintexts with $K$ groups, $18K+8$ random variables will be randomly generated at the encryption party. After encryption is completed, ciphertext text with $19K+8$ groups  will be generated. After encryption is completed, the encryption party will delete these generated intermediate random variables. For the attacker, knowing the plaintext-ciphertext correspondences, the problem of cracking the key is equivalent to solving a system of equations. In this chapter, we will prove that the computational complexity of cracking the key is exponential to the length of the key, this also means that decryption is a one-way function, thus solving the P vs NP problem based on the conclusion that ``one-way functions exist means P$\neq$NP".\\
\indent
In fact the encryption algorithm developed in Chapter 5 does not meet the security requirements for the key. This is because, when knowing the  plaintext-ciphertext correspondence, although the complexity of cracking the key is exponential to the length of the key, it does not guarantee that only exhaustive search is effective for cracking the key. There may be some other attack methods that can effectively reduce the difficulty of cracking the key. Such as for a 768 bit key, although it can ensure that the computational complexity of cracking the key is at least $O(2^{64})$, this is still far from the ideal security of $O(2^{768})$. The encryption algorithm constructed in Chapter 5 only analyzes the computational complexity of cracking the key given plaintext ciphertext correspondences, which does not analyzes its ability to resist linear and differential attacks, For attackers, they can construct plaintexts with certain characteristics multiple times, then find some structures that we did not anticipate from the relationship between plaintext and ciphertext.\\
\indent
In Chapter 6, we first construct nonlinear operations of one parameter on another, such as ``high-order" operations. Based on this nonlinear construction, we redesigned a block symmetric encryption algorithm and named it $Eagle^{*}$. For  $Eagle^{*}$, knowing any plaintext ciphertext correspondence, there is no more effective way for attackers to crack the key than exhaustive search. This is because the key is consist of three variables, the problem of verifying any two variables in the key is equivalent to solving a system of quintic equations. Based on the conclusion that there are no algebraic solutions to quintic equations, we assert that there is no fast algorithm to verify any two variables, and thus the computational complexity of cracking the key is equal to the computational complexity of completely exhausting the three variables.\\
\indent
On the other hand, for the $Eagle^{*}$ encryption algorithm, The ciphertext can be seen as four independent parts. Three parts of the ciphertext is completely determined by the intermediate random number which is completely unrelated to the plaintext. The fourth part of the ciphertext is determined by the intermediate random number together with the plaintext sequence, but it is completely indistinguishable whether the uncertainty of the ciphertext brought by the random number or brought by different plaintexts. This theoretically ensure that $Eagle^{*}$ can resist all forms of linear attacks and differential attacks.\\

\section{Notations}
\indent
Some notations \\
\indent
$\oplus$  - XOR operation.\\
\indent
$D^{+n}$  - Do left cycle shift of D by n bits. \\
\indent
$m[i]$  - The $i-$th bit of $m$. \\
\indent
$\mathbb{E}_{w_{0}, w_{1}}(m)$  - Excute Eagle encoding algorithm as shown in Algorithm 1.\\
\indent
$\mathbb{D}_{w_{0}, w_{1}}(s)$  - Excute Eagle decoding algorithm as shown in Algorithm 2. \\
\indent
$m * \delta$  - Excute the right multiplication of $m$ with $\delta$ as shown in Algorithm 3. \\
\indent
$s /  \delta$ - Excute the right division of $s$ to $\delta$ as shown in Algorithm 4. \\
\indent
$m * \delta^{2}$ - Excute the right multiplication of  $m$ with $\delta$ twice as $m * \delta * \delta$ as shown in Algorithm 9.  \\
\indent
$s /  \delta^{2}$ - Excute the right division of $s$ to $\delta$ twice as $s / \delta / \delta$ as shown in Algorithm 10. \\
\indent
$(m_{1}, m_{2}) \bigotimes (k_{1}, k_{2}) = (c_{1}, c_{2})$ - A binary operation of two-dimensional vectors defined in chapter 4.2. \\
\indent
$\phi(x)$ - A simple function changes $x$ to an odd number if $x$ contains odd number of bits that are 1, as shown in Algorithm 5. \\
\indent
$\psi(x)$ - A simple function changes $x$ to a number contains odd number of bits that are 1 if $x$ is an odd number, as shown in Algorithm 6. \\

\section{Introduction to Eagle encoding algorithm}
\indent
We first introduce two common bit operations. XOR denoted as  $\oplus$. Do left cycle shift of $D$ by $n$ bits which can be denoted as $D^{+n}$, for example
$(10011010)^{+2}=(01101010)$ .\\
\indent
We select two $L$-bits parameters  $w_{0}$ and $w_{1}$, have odd number of different bits. For example $w_{0}=10010011$ and $w_{1}=11000111$ have 3 bits (3 is odd) different.\\
\begin{equation}
\begin{aligned}
w_{0} = 1 \textcolor{red}{0} 0 \textcolor{red}{1} 0 \textcolor{red}{0} 1 1 \\\notag
w_{1} = 1 \textcolor{red}{1} 0 \textcolor{red}{0} 0 \textcolor{red}{1} 1 1 \notag
\end{aligned}
\end{equation}
\indent
We set the initial state of  $L$-bit as $s_{0}$, we choose one parameter $w \in \{w_{0},w_{1}\}$, without loss of generality, assume that we choose $w=w_{0}$, then we define the following calculation\\
\begin{equation}
s_{1}=w\oplus(s_{0}\oplus s_{0}^{+1})=w_{0}\oplus(s_{0}\oplus s_{0}^{+1})  \tag{3.1}
\end{equation}
\indent
From (3.1), we can easily know\\
\begin{equation}
s_{0}\oplus s_{0}^{+1}  = s_{1}\oplus w \tag{3.2}                                                                                                                                                                                            
\end{equation}
\indent
If we only know  $s_{1}$ but not $s_{0}$, we don ’t know whether $w=w_{0}$ or $w=w_{1}$  we used in (3.1), we can check it through a simple trial-and-error test. For example, we guess the parameter $w=w_{1}$ was used in (3.1), we need to find $s_{x}$ to satisfy\\
\begin{equation}
s_{x} \oplus s_{x}^{+1}  = s_{1} \oplus w_{1} \tag{3.3}
\end{equation}
\indent
In fact, since  $w_{0}$ and $w_{1}$ have odd number of different bits, such $s_{x}$ does not exist. See Theorem 1 for details.\\\newline
\indent
\textbf{[Theorem 1]} Two  $L$-bit parameters $w_{0}$ and $w_{1}$ which have odd number of different bits, given $s_{0}$, we set $s_{1}= w_{0} \oplus (s_{0} \oplus s_{0}^{+1})$ , then there doesn’t exists $s_{x}$ satisfy  $s_{x} \oplus s_{x}^{+1} = s_{1} \oplus w_{1}$.\\
\begin{proof}
\indent
Firstly, by definition we have\\
\begin{equation}
s_{1} \oplus w_{1} = w_{0} \oplus (s_{0} \oplus s_{0}^{+1}) \oplus w_{1}  = (s_{0} \oplus s_{0}^{+1}) \oplus (w_{0} \oplus w_{1}) \tag{3.4}
\end{equation}
\indent
Where $w_{0}$ and $w_{1}$ have odd number of different bits, so there are odd number of 1 in the bit string of $w_{0} \oplus w_{1}$ .\\
\indent
Proof by contradiction, we suppose that there exists $s_{x}$  satisfy $s_{x} \oplus s_{x}^{+1} = s_{1} \oplus w_{1}$, then by (3.4), we have\\
\begin{equation}
s_{x} \oplus s_{x}^{+1} = (s_{0} \oplus s_{0}^{+1}) \oplus (w_{0} \oplus w_{1})  \tag{3.5}
\end{equation}
\indent
By simple calculation we have\\
\begin{equation}
(s_{0} \oplus s_{x}) \oplus (s_{0} \oplus s_{x})^{+1} = w_{0} \oplus w_{1}  \tag{3.6}
\end{equation}
\indent
We set  $s_{y} = s_{0} \oplus s_{x}$, then there are odd number 
of 1 in the bit string of  $s_{y} \oplus s_{y}^{+1}$, without of generality, we suppose that the bits with 1 are  $l_{1}, l_{2}, ..., l_{u}$ ($u$  is odd). Compare to the first bit of $s_{y}$, the $l_{1}+1$ bit of $s_{y}$ is different from the first bit of $s_{y}$  , the $l_{2}+1$ bit of $s_{y}$ is the same with the first bit of  $s_{y}$, the $l_{3}+1$ bit of $s_{y}$ is different from the first bit of $s_{y}$, the $l_{4}+1$ bit of $s_{y}$ is the same with the first bit of $s_{y}$, and so on, since $u$ is odd, $u-1$ is even, the $l_{u-1}+1$ bit of $s_{y}$ is the same with the first bit of $s_{y}$. \\
\indent
If $l_{u}<L$, the $l_{u}+1$ bit of $s_{y}$ is different from the first bit of $s_{y}$, so the last bit of $s_{y}$ is different from the first bit of $s_{y}$. On the other hand, the last bit of $s_{y} \oplus s_{y}^{+1}$ is 0, so the last bit of $s_{y}$ is the same with the first bit of $s_{y}$. This is contradictory. \\
\indent
\begin{tikzcd}
a \arrow[rd] \arrow[d] & b \arrow[d] \arrow[rd] & b \arrow[d] \arrow[rd] & a \arrow[d] \arrow[rd] & a \arrow[d] \arrow[rd] & b \arrow[d] \arrow[rd] & b \arrow[d] \arrow[rd] & b \arrow[d] \arrow[llllllld] \\
0                      & 1 \arrow[d, dotted]    & 0                      & 1 \arrow[d, dotted]    & 0                      & 1 \arrow[d, dotted]    & 0                      & 0                            \\
                       & l_{1}                  &                        & l_{2}                  &                        & l_{3}                  &                        &                             
\end{tikzcd}
\\
\indent
If $l_{u}=L$, the last bit of $s_{y}$ is different from the the first bit of $s_{y}$. On the other hand, the last bit of $s_{y}$ is the same with $l_{u-1}$ bit of $s_{y}$ which is the same with the first bit of $s_{y}$. This is contradictory. \\
\indent
\begin{tikzcd}
a \arrow[rd] \arrow[d] & b \arrow[d] \arrow[rd] & b \arrow[d] \arrow[rd] & a \arrow[d] \arrow[rd] & a \arrow[d] \arrow[rd] & a \arrow[d] \arrow[rd] & a \arrow[d] \arrow[rd] & b \arrow[d] \arrow[llllllld] \\
0                      & 1 \arrow[d, dotted]    & 0                      & 1 \arrow[d, dotted]    & 0                      & 0                      & 0                      & 1 \arrow[d, dotted]          \\
                       & l_{1}                  &                        & l_{2}                  &                        &                        &                        & l_{3}                       
\end{tikzcd}
\\ 
\indent
So there doesn’t exists $s_{x}$ satisfy $s_{x} \oplus s_{x}^{+1} = s_{1} \oplus w_{1}$ .
\end{proof}
\indent
Let's go back to the discussion just now, after a trial-and-error test, we can accurately check which one ( $w=w_{0}$ or $w=w_{1}$ ) we just used in (3.1).\\
\indent
Now we suppose that there is a binary sequence $m = b_{1}b_{2} ...b_{L}$  with length $L$. Start with $s_{0}$,  read each bit of $M$ from left to right  sequentially,  when the bit $b[i](1 \leq i \leq L)$ is 0, we set $s_{i} = w_{0} \oplus (s_{i-1} \oplus s_{i-1}^{+1})$, when the bit  $b[i]$ is 1, we set $s_{i} = w_{1} \oplus (s_{i-1} \oplus s_{i-1}^{+1})$.\\
\indent
According to the above calculation, for every $s_{i}$, we can find the only  $w_{x}=w_{0}$ or $w_{x}=w_{1}$
such that there exists $s_{i-1}$ satisfy $s_{i -1} \oplus s_{i -1}^{+1} = s_{i} \oplus w_{x}$.  According to the properties of XOR and cyclic shift, we can easily see that there are only two $s_{i-1}$ that satisfy $s_{i-1} \oplus s_{i-1}^{+1} = s_{i} \oplus w_{x}$, and the two $s_{i-1}$ with each bit different. As long as we know any one bit of $s_{i-1}$, $s_{i-1}$ can be uniquely determined. So we only need to save one bit of $s_{i}$, finally by $s_{L}$, we can completely restore the original state $s_{0}$ and the binary sequence $m$.\\
\indent
Based on the above discussion, we can construct a complete Eagle encoding and decoding algorithm. The entire algorithm consists of three processes: generating parameters, encoding, and decoding.\\\newline
\indent
\textbf{[Parameter generation]}\\
\indent
Firstly we choose two  $L$-bit parameters $w_{0}$ and  $w_{1}$ which have odd number of different bits, then we choose $L$-bit initial state $s_{0}$.\\
\\
\indent
\textbf{[Encoding]}\\
\indent
For input data $m$, we record $m[i](1\leq i \leq L)$ as the $i$-th bit of  $m$, $m[i] \in \{0, 1\}$, $L$ is the length of $m$, the encoding process is as follows.\\

\begin{algorithm} [H]
\caption{$Eagle$-Encoding.}
\label{alg: Framework}
\begin{algorithmic}[1]
\REQUIRE ~~ \\
Parameter: $(w_{0}, w_{1})$. \\
Input: $m=[m[1], m[2], ..., m[L]]$. \\
Random number: $s_{0}$. \\
\ENSURE ~~ \\
Output: $(c, s_{L})$, \\
\STATE Define $c$ as a bit array with length $L$. \\
\FOR{$i=1; i \le L; i++$}
	\STATE $c[i] \leftarrow s_{i-1}[L]$ \\
	\IF{m[i] = 0}
		\STATE $s_{i} \leftarrow  w_{0} \oplus (s_{i-1} \oplus s_{i-1}^{+1})$ \\
	\ELSE
		\STATE $s_{i} \leftarrow  w_{1} \oplus (s_{i-1} \oplus s_{i-1}^{+1})$ \\
	\ENDIF
\ENDFOR
\end{algorithmic}
\end{algorithm}
\indent
Where the $i$-th bit of $c$ takes the last bit of state $s_{i-1}$. \\

\indent
\textbf{[Decoding]}\\
\indent
The output $(c , s_{L})$ of the above  encoding process is used as the input of the decoding process, the decoding process is as follows. \\

\begin{algorithm} [H]
\caption{$Eagle$-Decoding.}
\label{alg: Framework}
\begin{algorithmic}[1]
\REQUIRE ~~ \\
Parameter: $(w_{0}, w_{1})$. \\
Input: $c=[c[1], c[2], ..., c[L]], s_{L}$. \\
\ENSURE ~~ \\
Output: $m$. \\
\FOR{$i=L; i \ge 1; i--$}
	\STATE Define $x$ as a bit. $(x = 0 \, or \, x = 1)$ \\
	\STATE Do trial-and-error testing with  $s_{i} \oplus w_{0}$ or $s_{i} \oplus w_{1}$ , find the unique $w_{x} ( x = 0 \, or \, x = 1 )$ satisfy $s_x \oplus s_{x}^{+1} = s_{i} \oplus w_{x}$.\\
	\STATE $m[i] \leftarrow x$ \\
	\STATE For the two possible $s_{x}$ satisfy $s_{x} \oplus s_{x}^{+1} = s_{i} \oplus w_{x}$, we select the one which the last bit is equal to $c[i]$ as $s_{i -1}$. \\
\ENDFOR
\end{algorithmic}
\end{algorithm}

\indent
Now we give an example of the above algorithms.\\
\indent
\textbf{[Example 1] Eagle encoding} \\
\indent
We choose the parameters as $w_{0}=10010011$, $w_{1}=11000111$, $s_{0}=01011001$, $m=10010101$.\\
\indent
Since $m[1]=1$, we set $w=w_{1}$, then we have
\begin{equation}
s_{1} = w \oplus (s_{0} \oplus s_{0}^{+1})=11000111 \oplus 11101011 = 00101100 \notag
\end{equation}
\indent
Since $m[2]=0$, we set $w=w_{0}$, then we have
\begin{equation}
s_{2} = w \oplus (s_{1} \oplus s_{1}^{+1})=10010011 \oplus 01110100 = 11100111 \notag
\end{equation}
\indent
Since $m[3]=0$, we set $w=w_{0}$, then we have
\begin{equation}
s_{3} = w \oplus (s_{2} \oplus s_{2}^{+1})=10010011 \oplus 00101000 = 10111011 \notag
\end{equation}
\indent
Since $m[4]=1$, we set $w=w_{1}$, then we have
\begin{equation}
s_{4} = w \oplus (s_{3} \oplus s_{3}^{+1})=11000111 \oplus 11001100 = 00001011 \notag
\end{equation}
\indent
Since $m[5]=0$, we set $w=w_{0}$, then we have
\begin{equation}
s_{5} = w \oplus (s_{4} \oplus s_{4}^{+1})=10010011 \oplus 00011101 = 10001110 \notag
\end{equation}
\indent
Since $m[6]=1$, we set $w=w_{1}$, then we have
\begin{equation}
s_{6} = w \oplus (s_{5} \oplus s_{5}^{+1})=11000111 \oplus 10010011 = 01010100 \notag
\end{equation}
\indent
Since $m[7]=0$, we set $w=w_{0}$, then we have
\begin{equation}
s_{7} = w \oplus (s_{6} \oplus s_{6}^{+1})=10010011 \oplus 11111100 = 01101111 \notag
\end{equation}
\indent
Since $m[8]=1$, we set $w=w_{1}$, then we have
\begin{equation}
s_{8} = w \oplus (s_{7} \oplus s_{7}^{+1})=11000111 \oplus 10110001 = 01110110 \notag
\end{equation}
\indent
Take the last bit of $s_{0},s_{1},s_{2},s_{3},...,s_{7}$ as $c= 10111001$. \\
\indent
We use $(c, s_{8})=(10111001, 01110110)$ as the results generated by Eagle encoding. \\
\indent
Next, we will provide the example of Eagle decoding.\\\newline
\indent
\textbf{[Example 2] Eagle decoding} \\
\indent
Known $w_{0}=10010011$, $w_{1}=11000111$, $s_{8}=01110110$, $c=10111001$, how to compute $m$?\\
\indent
Firstly by $s_{8} = w \oplus (s_{7}  \oplus s_{7}^{+1})   =>  s_{7} \oplus s_{7}^{+1} = s8 \oplus w$, we don't know whether $w=w_{0}$ (the 8-th bit of $m$ is 0) or $w=w_{1}$ (the 8-th bit of $m$ is 1), we need a trial-and-error test.  \\
\indent
Let's assume $w = w_{0}$, we have
\begin{equation}
s_{7} \oplus s_{7}^{+1} = s_{8} \oplus w_{0} = 01110110 \oplus 10010011 = 11100101 \notag
\end{equation}  
\indent
There is no such $s_{7}$ satisfies $s_{7} \oplus s_{7}^{+1} = 11100101$. \\
\indent
\begin{tikzcd}
a \arrow[dd] \arrow[rrrrrrrdd] & b \arrow[ldd] \arrow[dd] & a \arrow[ldd] \arrow[dd] & b \arrow[ldd] \arrow[dd] & b \arrow[ldd] \arrow[dd] & b \arrow[ldd] \arrow[dd] & a \arrow[ldd] \arrow[dd]          & a \arrow[dd] \arrow[ldd] \\
                               &                          &                          &                          &                          &                          &                                   &                          \\
1                              & 1                        & 1                        & 0                        & 0                        & 1                        & 0                                 & 1 \arrow[ld, dotted]     \\
                               &                          &                          &                          &                          &                          & {a \oplus a=1} &                         
\end{tikzcd}
\\\newline
\indent
Let's assume $w=w_{1}$, we have
\begin{equation}
s_{7} \oplus s_{7}^{+1} = s_{8} \oplus w_{1} = 01110110 \oplus 11000111 = 10110001 \notag
\end{equation}
\indent
There exists two solutions $s_{7}= 01101111$ and $ s_{7} = 10010000$, these two solutions are bit reversed.\\
\indent
Since the 8-th bit of $c$ is $c[8]=1$, the last bit of $s_{7}$ is 1, we have $s_{7} = 01101111$.\\
\indent
According to the above, because $w=w_{1}$, the 8-bit of $m$ is 1.\\
\indent
Similarily we get $m=10010101$. \\\newline
\indent
It is not difficult to find that the above encoding process and decoding process are correct, that  is  $(c , s_{L})$ generated  by encoding from $m$ can  be  completely restored  through  the  decoding   process. In  addition,  the  encoding   process  is sequential  encoding in the order of $m$’s bits,  and  the  decoding  process  is sequential decoding in the reverse order of  $c$’s bits.\\
\indent
\begin{tikzcd}
s_{0} \arrow[r, "{m[1]}"] & s_{1} \arrow[r, "{m[2]}"] & s_{2} \arrow[r, "{m[3]}"] & ... \arrow[r, "{m[L]}"] & s_{L}                    \\
s_{0}                     & s_{1} \arrow[l, "{c[1]}"]  & s_{2} \arrow[l, "{c[2]}"]  & ... \arrow[l, "{c[3]}"]  & s_{L} \arrow[l, "{c[L]}"]
\end{tikzcd}
\\
\indent
We also noticed the fact that  in the above encoding and decoding  process, all inputs and outputs do not need to appear $s_{0}$ , This means that the selection of $s_{0}$  will not affect the correctness of the encoding process and decoding process. The arbitrary of $s_{0}$ will bring the unpredictability of the encoded output.\\
\indent
For the convenience of the discussion in the following chapters, here we briefly analyze the effect of unpredictability of $s_{0}$ on the encoded output.\\
\indent
Given  the parameters  $w_{0}$ and $w_{1}$  that have odd number of different bits, for a certain input $m$   of  $L$   bits, since   $s_{0}$      is arbitrarily selected, it is obvious that $c$ is uncertain, but is the final state $s_{L}$ necessarily uncertain?\\
\indent
In fact, the answer is no. In some cases, such as  $L = 2^{u}$  (that is, the parameter length is the power of 2), the final state $s_{L}$ is determined for different choices of $s_{0}$. The final state variable $s_{L}$  which is the output of the encoding process is only related to the input  $m$  and has nothing to do with the choice of the initial state $s_{0}$. See Theorem 2 for details.\\\newline
\indent
\textbf{[Theorem 2]} In Eagle encoding algorithm, given the parameters $w_{0}$ and $w_{1}$  that have $L$  bits with odd bits different, for a certain $L$  bit  input $m$, if $L = 2^{u}$ is satisfied, then for any initial state $s_{0}$, after the Eagle encoding process, the final state  $s_{L}$ is only related to the input $m$, and is unrelated with the choice of the initial state $s_{0}$.\\
\indent
\begin{proof}
We represent  $m$ as binary stream  $x_{1}x_{2} ...x_{L}$ , which  $x_{i} \in \{0, 1\}$, $1 \leq i \leq L$.

We execute the Eagle encoding process to  $m$  from  $x_{1}$     to  $x_{L}$     as follows.\\
\begin{equation}
s_{1}  = w_{x_{1}} \oplus (s_{0} \oplus s_{0}^{+1}) = f_{1} (w_{x_{1}} ) \oplus (s_{0}  \oplus s_{0}^{+1}) \notag
\end{equation}
\begin{equation}
s_{2}  = w_{x_{2}} \oplus (s_{1} \oplus s_{1}^{+1}) = f_{2} (w_{x_{1}} , w_{x_{2}} ) \oplus (s_{0}  \oplus s_{0}^{+2}) \notag
\end{equation}
\begin{equation}
s_{3}  = w_{x_{3}} \oplus (s_{2} \oplus s_{2}^{+1}) = f_{3} (w_{x_{1}}  , w_{x_{2}}  , w_{x_{3}}) \oplus (s_{0} \oplus s_{0}^{+1} \oplus s_{0}^{+2} \oplus s_{0}^{+3})  \notag
\end{equation}
\begin{equation}
s_{4}  = w_{x_{4}} \oplus (s_{3} \oplus s_{3}^{+1}) = f_{4} (w_{x_{1}}  , w_{x_{2}}  , w_{x_{3}}  , w_{x_{4}}  ) \oplus (s_{0}  \oplus s_{0}^{+4}) \notag
\end{equation}
\indent
It  is  not  difficult  to  find  that  for  any    $m = 2^{v}$,   $s_{m}   = f_{m} (w_{x_{1}}  , ..., w_{x_{m}}  ) \oplus (s_{0} \oplus s_{0}^{+m} )$  holds,  this   can  be  proved  by  a  simple  mathematical  induction. \\
\indent
In  fact,  the conclusion is correct for  $v = 1$. \\
\indent
We assume that the conclusion is correct for  $v -1$ , we  have \\
\begin{equation}
s_{m/2}  = f_{m/2} (w_{x_{1}}  , ..., w_{x_{m/2}}  ) \oplus (s_{0} \oplus s_{0}^{+m/2} ) \notag
\end{equation}
\indent
Since    $s_{m/2}$         to   $s_{m}$          must  do calculations with  $m / 2$  steps, we have\\
\begin{equation}
\begin{aligned}
s_{m} = f_{m} (w_{x_{1}}  , ..., w_{x_{m}}  ) \oplus (s_{0}  \oplus s_{0}^{+m/2} ) \oplus (s_{0}^{+m/2}  \oplus s_{0}^{+m} ) \\= f_{m} (w_{x_{1}}  , ..., w_{x_{m}}  ) \oplus (s_{0}  \oplus s_{0}^{+m} )       \notag
\end{aligned}
\end{equation}
\indent
Since  $L = 2^{u}$, we have  $s_{L}  = f_{L} (w_{x_{1}}  , ..., w_{x_{L}}  ) \oplus (s_{0}  \oplus s_{0}^{+L} )$ , where  $f_{i} (...)$   is irrelevant with  $s_{0}$, by definition of cycle shift, we have  $s_{0}  = s_{0}^{+L}$, so  $s_{L}  = f_{L} (w_{x_{1}}  , ..., w_{x_{L}}  )$ which  is irrelevant with  $s_{0}$.\\\newline
\end{proof}
\indent
From theorem 2, for any parameters   $w_{0}$     and  $w_{1}$    with length  $L = 2^{u}$, for any initial state  $s_{0}$, execute Eagle encoding on  $m$   to obtain  $s_{L}$     which is irrelevant with $s_{0}$. \\
\indent
Next we introduce two symbols  $\mathbb{E}$ and $\mathbb{D}$.\\
\indent
$\mathbb{E}_{w_{0},w_{1}}(m)=s_{L}$ : use the key $(w_{0} , w_{1})$  to execute Eagle encoding algorithm (Algorithm 1) on any initial state  $s_{0}$  and the given input  $m$   to obtain  $c$  and  $s_{L}$, which $s_{0}$ and $c$ will not be analyzed anymore. \\
\indent
$\mathbb{D}_{w_{0},w_{1}}(s_{L})=m$ : use the key $(w_{0} , w_{1})$    to execute Eagle decoding algorithm (Algorithm 2) on any $c$  and the given $s_{L}$    to obtain  $s_{0}$  and  $m$.\\
\indent
$\mathbb{E}$ and $\mathbb{D}$ both represent a complete Eagle encoding process and Eagle decoding process, and their introduction is mainly for the convenience of deriving encryption algorithms later.  \\
\indent
In all the following chapters of this paper, we assume the length $L$ is a power of 2.\\

\section{In depth analysis of Eagle encoding algorithm}
\indent
Back to the Eagle encoding algorithm, we noticed that $w_ {0}$ differs from $w_ {1}$ with an odd number of bits, which is equivalent to the binary representation of $w_ {0} \oplus w_ {1}$ containing an odd number of 1. There are $2^{2L-1}$ pairs of $(w_ {0}, w_ {1})$ that satisfy an odd number of different bits, This is because among all $(w_ {0}, w_ {1})$, $w_ {0}$ and $w_ {1}$ contain the same number of odd bits different as even bits different. \\
\indent
Next, let's take a closer look at the features of the Eagle encoding algorithm. For the parameter $(w_ {0}, w_ {1})$, we have another completely equivalent representation $(\delta, w_ {0})$, where $\delta=w_ {0} \oplus w_ {1}$, which means that $(w_ {0}, w_ {1})$ and $(\delta, w_ {0})$ can be equivalently transformed. Since the selection of $s_ {0}$ is independent of $s_ {L}$, without loss of generality, we set $s_{0}=0$, when given $(\delta, w_ {0})$, we only observe the relationship between the final state $s_ {L}$ and the input $m=b_ {1}b_ {2}...b_{L}$, where $b_{i} \in \{0,1\}$. We excute the Eagle encoding algorithm with the first bit to obtain \\
\begin{equation}
s_{1}=w_{b_{1}} \oplus (s_{0} \oplus s_{0}^{+1}) = w_{0} \oplus g_{1}(b_{1}, \delta) \notag
\end{equation}
\indent
Where $g_{1}(b_{1}, \delta)=0$ when $b_{1}=0$ and $g_{1}(b_{1}, \delta)=\delta$ when $b_{1}=1$. We excute the Eagle encoding algorithm with the second bit to obtain \\
\begin{equation}
\notag
\begin{aligned}
s_{2}=w_{b_{2}} \oplus (s_{1} \oplus s_{1}^{+1}) \\
 =(w_{0} \oplus (w_{0} \oplus w_{0}^{+1})) \oplus g_{2}(b_{1}, b_{2}, \delta) \\
 =w_{0}^{+1} \oplus g_{2}(b_{1}, b_{2}, \delta) 
\end{aligned}
\end{equation}
\indent
Where $g_{2}(b_{1}, b_{2}, \delta)$ is a function about $\delta$ when $b_{1}, b_{2}$ is given. We excute the Eagle encoding algorithm with the third  bit to obtain \\
\begin{equation}
\notag
\begin{aligned}
s_{3}=w_{b_{3}} \oplus (s_{2} \oplus s_{2}^{+1}) \\
=(w_{0} \oplus w_{0}^{+1} \oplus w_{0}^{+2}) \oplus g_3(b_{1}, b_{2}, b_{3}, \delta)
\end{aligned}
\end{equation}
\indent
So far we see that $s_{i}=h_{i}(w_{0}) \oplus g_{i}(b_{1}, b_{2}, ..., b_{i}, \delta)$, where $h_{i}(w_ {0})$ can be regarded as the state obtained by executing $i$ bits Eagle encoding algorithm when $m=0$, $g_{i}(b_{1}, b_{2}, ..., b_{i}, \delta)$ is a function about $\delta$ when $b_{1}, b_{2}, ..., b_{i}$ are given. \\
\indent
When $L=2^{u}$, we set $m=0, s_{0}=0$, after executing Eagle encoding, the final state is $s_{L}=w_{0}^{+(L-1)}$. When $m \neq 0$, after excuting Eagle encoding, the final state is \\
\begin{equation}
s_{L}=w_{0}^{+(L-1)} \oplus g_{L}(b_{1}, b_{2}, ..., b_{L}, \delta) \tag{4.1}
\end{equation}
\indent
Where $g_{L}(b_{1}, b_{2}, ..., b_{L}, \delta)$ can also be expressed as $g_{L}(m, \delta)$. It is not difficult to see that $g_{L}(m, \delta)$ is a series of XOR operations with $\delta^{+k}$ given $m$, this is also equivalent to $g_{L}(m, \delta)=c_{1}\delta^{+1} \oplus c_{2} \delta^{+2} \oplus ... \oplus c_{L} \delta^{+L}$, where $c_{i} \in \{0, 1\}$ and $(c_{1},c_{2},...,c_{L})$ only determined by $m$. So (4.1) can ultimately be expressed as \\
\begin{equation}
s_{L}=w_{0}^{+(L-1)} \oplus g_{L}(m, \delta)=w_{0}^{+(L-1)} \oplus (c_{1}\delta^{+1} \oplus c_{2} \delta^{+2} \oplus ... \oplus c_{L} \delta^{+L}) \tag{4.2}
\end{equation}
\indent
Furthermore, we observe several important facts, as shown in the following theorems.\\
\\
\indent
\textbf{[Theroem 3]} $\forall m,s$, there exists $2^{L-1}$ different groups $(w_{0},w_{1})$ satisfy that $\mathbb{E}_{w_{0},w_{1}}(m)=s$. For any $\delta$, there exists a unique $w_{0}$ that $\mathbb{E}_{w_{0},\delta \oplus w_{0}}(m)=s$. \\
\indent
\begin{proof}
\indent
From (4.2), we know $\mathbb{E}_{w_{0},w_{1}}(m)=w_{0}^{+(L-1)} \oplus g_{L}(m, \delta)$, through a simple transformation we obtain $w_{0}^{+(L-1)}=g_{L}(m, \delta) \oplus s$, it is obvious that for any $\delta, m, s$, the right side of the equation is a difinite result, so there exists a unique $w_{0}$ valid. \\
\indent
Additionally, due to the odd number of bits differ between $w_ {0}$ and $w_ {1}$, obviously $\delta=w_{0} \oplus w_{1}$ have an odd number of bits that are 1, that is to say, there are $2^{L-1}$ different $\delta$. \\
\end{proof}
\indent
The conclusion of Theorem 3 indicates that, for any $\mathbb{E}_{w_{0}, w_{1}}(m)=s$, when only given $(m, s)$, there are exponentially different $(w_{0}, w_{1})$ valid. \\
\\
\indent
\textbf{[Theorem 4]} For any $m_{1}, s_{1}, m_{2}, s_{2}$, when the parity between $m_{1}$ and $m_{2}$ is different, there exists a unique $(\delta, w_{0})$ that $\mathbb{E}_{w_{0},\delta \oplus w_{0}}(m_{1})=s_{1}$ and $\mathbb{E}_{w_{0},\delta \oplus w_{0}}(m_{2})=s_{2}$. \\
\begin{proof}
\indent
The state obtained by excuting the Eagle encoding algorithm with $L-1$ bits on $m_{1}$ is $s_{L-1,1}=h_{L-1,1}(w_{0}) \oplus g_{L-1,1}(m_{1}, \delta)$,  the state obtained by excuting the Eagle encoding algorithm with $L-1$ bits on $m_{2}$ is $s_{L-1,2}=h_{L-1,2}(w_{0}) \oplus g_{L-1,2}(m_{2}, \delta)$. \\
\indent
Due to the difference in parity between $m_{1}$ and $m_{2}$, it can be assumed that the last bit of $m_{1}$ is 0 and the last bit of $m_{2}$ is 1. Further execute the Eagle encoding algorithm with one more bit separately to obtain \\
\begin{equation}
\notag
\begin{aligned}
s_{1}=s_{L,1}=w_{0}^{+(L-1)} \oplus (g_{L-1,1}(m_{1}, \delta) \oplus g_{L-1,1}(m_{1}, \delta)^{+1}) \\
s_{2}=s_{L,2}=w_{0}^{+(L-1)} \oplus \delta \oplus (g_{L-1,2}(m_{2}, \delta) \oplus g_{L-1,2}(m_{2}, \delta)^{+1})
\end{aligned}
\end{equation}
\indent
Perform XOR operation on both sides of the above two equations simultaneously, set $(g_{L-1,1}(m_{1}, \delta) \oplus (g_{L-1,2}(m_{2}, \delta)$ as $g( m_{1}, m_{2}, \delta)$ to obtain \\
\begin{equation}
s_{1} \oplus s_{2}=\delta \oplus (g(m_{1}, m_{2}, \delta) \oplus g(m_{1}, m_{2}, \delta)^{+1}) \notag
\end{equation}
\indent
Obviously, the left side of the equation is a definite number, on the right side of the equation is the result of performing XOR operation on an odd number of $\delta^{+k}$, It is not difficult to verify that the solution to such an equation exists and is unique. \\
\end{proof}
\indent
The conclusion of Theorem 4 indicates that, for any $\mathbb{E}_{w_{0}, w_{1}}(m_{1})=s_{1}$ and $\mathbb{E}_{w_{0}, w_{1}}(m_{2})=s_{2}$, when only given $m_{1}, s_{1}, m_{2}, s_{2}$, there is a 50\% probability that a unique $w_{0}, w_{1}$ valid. \\
\\
\indent
\textbf{[Theorem 5]} For any two groups $m_{1}, m_{2}$ and $m_{1}^{'}, m_{2}^{'}$, if $m_{1} \oplus m_{2}=m_{1}^{'} \oplus m_{2}^{'}$, excute the Eagle encoding algorithm as (4.2) on $m_{1}, m_{2}, m_{1}^{'}, m_{2}^{'}$ to obtain \\
\begin{equation}
\tag{4.3}
\begin{aligned}
\mathbb{E}_{w_{0}, \delta \oplus w_{0}}(m_{1})=w_{0}^{+(L-1)} \oplus g_{L}(m_{1}, \delta)  \\
\mathbb{E}_{w_{0}, \delta \oplus w_{0}}(m_{2})=w_{0}^{+(L-1)} \oplus g_{L}(m_{2}, \delta)  \\
\mathbb{E}_{w_{0}, \delta \oplus w_{0}}(m_{1}^{'})=w_{0}^{+(L-1)} \oplus g_{L}(m_{1}^{'}, \delta) \\
\mathbb{E}_{w_{0}, \delta \oplus w_{0}}(m_{2}^{'})=w_{0}^{+(L-1)} \oplus g_{L}(m_{2}^{'}, \delta) \\
\end{aligned}
\end{equation}
\indent
then $g_{L}(m_{1}, \delta) \oplus g_{L}(m_{2}, \delta)=g_{L}(m_{1}^{'}, \delta) \oplus g_{L}(m_{2}^{'}, \delta)$ holds.
\begin{proof}
\indent
Returning to the calculation process of (4.2), it is not difficult to find that \\
\begin{equation}
\mathbb{E}_{w_{0}, \delta \oplus w_{0}}(m \oplus \Delta m)=w_{0}^{+(L-1)} \oplus g_{L}(m, \delta) \oplus j(\delta, \Delta m) \notag
\end{equation}
\indent
Where $j(\delta, \Delta m)$ is a function only related to $\delta, \Delta m$. Without loss of generality we set $m_{1} \oplus m_{2}=m_{1}^{'} \oplus m_{2}^{'}=\Delta m$, (4.2) can be reorganized as \\
\begin{equation}
\tag{4.4}
\begin{aligned}
\mathbb{E}_{w_{0}, \delta \oplus w_{0}}(m_{1}^{'})=w_{0}^{+(L-1)} \oplus g_{L}(m_{1}, \delta) \oplus j(\delta, \Delta m) \\
\mathbb{E}_{w_{0}, \delta \oplus w_{0}}(m_{2}^{'})=w_{0}^{+(L-1)} \oplus g_{L}(m_{2}, \delta) \oplus j(\delta, \Delta m) \\
\end{aligned}
\end{equation}
\indent
By simple calculate we can obtain \\
\begin{equation}
\notag
\begin{aligned}
g_{L}(m_{1}^{'}, \delta) \oplus g_{L}(m_{2}^{'}, \delta) \\
=g_{L}(m_{1}, \delta) \oplus g_{L}(m_{2}, \delta) \oplus (w_{0}^{+(L-1)} \oplus w_{0}^{+(L-1)}) \oplus (j(\delta, \Delta m) \oplus j(\delta, \Delta m)) \\
=g_{L}(m_{1}, \delta) \oplus g_{L}(m_{2}, \delta)
\end{aligned}
\end{equation}
\end{proof}

\indent
Theorem 5 indicates that, the function $g_{L}(m, \delta)$ satisfies the distributive law for both variables $m$ and $\delta$. \\

\indent
\textbf{[Theorem 6]} For the function $g_{L}(m, \delta)$ in (4.2), fix any $\delta$, where $\delta$ contains an odd number of bits that are 1. Fix any $u$, there exists a unique $m$ that $g_{L}(m, \delta)=u$. \\
\begin{proof}
Fix any $\delta$, set $w_{0} \in \{0,1\}^{L}, w_{1}=w_{0} \oplus \delta$, excute Eagle encoding algorithm for $m$, we obtain $\mathbb{E}_{w_{0}, w_{1}}(m)=s$, then $u=s \oplus w_{0}^{+L}$, $g_{L}(m, \delta)=u$, thus $m \to u$ is monomorphism. \\
\indent
On the other hand, fix any $\delta$, set $w_{0} \in \{0,1\}^{L}, w_{1}=w_{0} \oplus \delta$, excute Eagle decoding algorithm for $s=u \oplus w_{0}^{+L}$, we obtain $\mathbb{D}_{w_{0}, w_{1}}(s)=m$, thus $u \to m$ is monomorphism. \\
\indent
Furthermore, since the ranges of $m$ and $u$ are the same, the mapping from $m$ to $s$ is a one-to-one corresponding when $\delta$ is fixed. \\
\end{proof}

\indent
The proof process of Theorem 6 actually describes the process of $g_{L}(m, \delta)=u$ and its inverse function $g_{L}^{-1}(u, \delta)$. \\

\indent
\subsection{The multiplication operation of binary}
\indent
According to the conclusion of Theorem 5, the symbol $g_{L}(m, \delta)$ satisfies the distributive law for both $m$ and $\delta$, that is \\
\begin{equation}
\notag
\begin{aligned}
g_{L}(m, \delta_{1} \oplus \delta_{2}) = g_{L}(m, \delta_{1}) \oplus g_{L}(m, \delta_{2}) \\
g_{L}(m_{1} \oplus m_{2}, \delta) = g_{L}(m_{1}, \delta) \oplus g_{L}(m_{2}, \delta)
\end{aligned}
\end{equation}
\indent
If we consider the XOR operation $\oplus$ as an addition operation to binary, $g_{L}(m, \delta)$ can be seen as the multiplication operation between $m$ and $\delta$, as follows \\
\begin{equation}
g_{L}(m, \delta) = m * \delta \notag
\end{equation}
\indent
It is not difficult to verify that $m * \delta$ does not satisfy the commutative law or the associative law, so $m * \delta$ can only be seen as the right multiplication of $\delta$ by $m$. It is obvious that this representation is well-defined. \\
\indent
The calculation of $m * \delta$ is shown in Algorithm 3 as follows. \\
\begin{algorithm} [H]
\caption{calculation of $m * \delta$.}
\label{alg: Framework}
\begin{algorithmic}[1]
\REQUIRE ~~ \\
Input: $m, \delta$. \\
\ENSURE ~~ \\
Output: $m * \delta$. \\
\STATE $w_{0} \leftarrow 0$. \\
\STATE $w_{1} \leftarrow \delta$. \\
\STATE $s_{L} \leftarrow \mathbb{E}_{w_{0}, w_{1}}(m)$. \\
\RETURN $s_{L}$. \\
\end{algorithmic}
\end{algorithm}
\indent
Similarly, we can also introduce the division operator $/$, using the notation $S/\delta=R$ to represent $R * \delta=S$. The calculation for division is as follows.\\
\begin{algorithm} [H]
\caption{calculation of $s / \delta$.}
\label{alg: Framework}
\begin{algorithmic}[1]
\REQUIRE ~~ \\
Input: $s, \delta$. \\
\ENSURE ~~ \\
Output: $s / \delta$. \\
\STATE $w_{0} \leftarrow 0$. \\
\STATE $w_{1} \leftarrow \delta$. \\
\STATE $m \leftarrow \mathbb{D}_{w_{0}, w_{1}}(s)$. \\
\RETURN $m$. \\
\end{algorithmic}
\end{algorithm}
\indent
This paper does not intend to delve into the algebraic properties of the right multiplication operation mentioned above, but only simplifies the representation of $g_{L}(m, \delta)$ using $m * \delta$. Therefore $\mathbb{E}_{w_{0}, w_{1}}(m)$ can be expressed as $m * (w_{0} \oplus w_{1}) \oplus w_{0}^{+(L-1)}$. \\
\indent
At the end of this section, we will provide the following conclusions.\\
\\
\indent
\textbf{[Theorem 7]} (i) If $m$ is an odd number, then $m * \delta$ contains odd number of bits that are 1. \\
\indent
(ii) If $m$ is an even number, then $m * \delta$ contains even number of bits that are 1. \\
\begin{proof}
\indent
Firstly we prove (i). \\
\indent
By definition, we have $m * \delta=\mathbb{E}_{0, \delta}(m)$. Let $s_{L-1}$ denote the result of the penultimate step of performing Eagle encoding on $m$, so we have $s_{L}=\delta \oplus (s_{L-1} \oplus s_{L-1}^{+1})$. \\
\indent
Obviously $(s_{L-1} \oplus s_{L-1}^{+1})$ contains even number of bits that are 1, $\delta$ contains odd number of bits that are 1, thus $s_{L}$ contains odd number of bits that are 1. \\
\indent
Similarly, (ii) also holds. \\
\end{proof} 
\indent
\textbf{[Theorem 8]} If $m$ is an odd number, $s$ contains odd number of bits that are 1, then there exists an unique $\delta$ that $m *  \delta = s$. \\
\indent
\begin{proof}
According to theorem 4, set $m_{1}=m, m_{2}=0, s_{1}=s, s_{2}=0$, because the parity between $m_{1}$ and $m_{2}$ is different, there exists an unique $(0, \delta)$ that $\mathbb{E}_{0, \delta}(m)=s$ which can be equivalently expressed as $m * \delta = s$. \\
\end{proof}

\subsection{Introduction of a binary operation of two-dimensional vectors}
\indent
We define a binary operation of two-dimensional vecotors $(m_{1}, m_{2}) \bigotimes (k_{1}, k_{2}) = (c_{1}, c_{2})$ as follows. \\
\begin{equation}
\tag{4.5}
\begin{aligned}
c_{2}=\mathbb{E}_{k_{1}, k_{2}}(m_{1}) \\
c_{1}=c_{2} \oplus (m_{2} * (k_{1} \oplus k_{2})) \\
\end{aligned}
\end{equation}
\indent
Where $k_{1}$ contains even number of bits that are 1, $k_{2}$ contains odd number of bits that are 1, $m_{1}$, $m_{2}$ are both odd numbers, $c_{1}$ contains even number of bits that are 1, $c_{2}$ contains odd number of bits that are 1. \\
\indent
The above operation can be represented by the following figure. \\
\\
\indent
\begin{tikzcd}
m_{1} \arrow[rd] \arrow[dd]                                   &                                     & m_{2} \arrow[ld] \arrow[dd]                    \\
                                                              & {k_{1},k_{2}} \arrow[ld] \arrow[rd] &                                                \\
{s_{1}=\mathbb{E}_{k_{1},k_{2}}(m_{1})} \arrow[rrd] \arrow[d] &                                     & s_{2}=m_{2} * (k_{1} \oplus k_{2}) \arrow[lld] \\
c_{1}=s_{1} \oplus s_{2}                                      &                                     & c_{2}=s_{1}                                   
\end{tikzcd}
\\
\\
\indent
It is not difficult to verify that the binary operation defined above satisfies the following conditions. \\
\indent
(a) Given any $(m_{1}, m_{2})$, the mapping $(k_{1}, k_{2}) \rightarrow (c_{1}, c_{2})$ is a bijection. \\
\indent
(b) Given any $(k_{1}, k_{2})$, the mapping $(m_{1}, m_{2}) \rightarrow (c_{1}, c_{2})$ is a bijection.\\
\indent
(c) Given any $(c_{1}, c_{2})$, the mapping $(k_{1}, k_{2}) \rightarrow (m_{1}, m_{2})$ is a bijection.\\
\indent
(d) Given only $m_{1}$ and $c_{1}$, any $(k_{1}, k_{2})$ can be used to solve for an unique $m_{2}$ and $c_{2}$. \\
\indent
(e) Given only $m_{2}$ and $c_{2}$, any $(k_{1}, k_{2})$ can be used to solve for an unique $m_{1}$ and $c_{1}$. \\
\indent
(f) Given only $m_{2}$ and $c_{1}$, any $(k_{1}, k_{2})$ can be used to solve for an unique $m_{1}$ and $c_{2}$. \\
\indent
(g) Given only $m_{1}$ and $c_{2}$, any $k_{1}$ can be used to solve for an unique $k_{2}$, any $k_{2}$ can also be used to solve for an unique $k_{1}$, any $k_{1} \oplus k_{2}$ can also be used to solve for an unique $k_{1}$ or $k_{2}$, and it is impossible to solve for $m_{2}$ or $c_{1}$ given any $k_{1}$ or $k_{2}$ or $k_{1} \oplus k_{2}$. \\
\indent
(h) Given only $m_{1}, c_{1}, c_{2}$, any $k_{1}$ can be used to solve for an unique $k_{2}, m_{2}$, any $k_{2}$ can also be used to solve for an unique $k_{1}, m_{2}$, any $k_{1} \oplus k_{2}$ can also be used to solve for an unique $k_{1}$ or $k_{2}$.  \\
\indent
(i) Given only $c_{1}, k_{1}, k_{2}$, any $m_{2}$ can be used to solve for an unique $c_{2}$, any $c_{2}$ can also be used to solve for an unique $m_{2}$.  \\
\indent
For condition (a), It is obvious that $(k_{1}, k_{2}) \rightarrow (c_{1}, c_{2})$ is an injective, but it is not so obvious that $(c_{1}, c_{2}) \rightarrow (k_{1}, k_{2})$ be also an injective. Since $c_{1} \oplus c_{2}=m_{2} * (k_{1} \oplus k_{2})$, from Theorem 8, we can solve an unique $k_{1} \oplus k_{2}$. By $c_{2}=\mathbb{E}_{k_{1}, k_{2}}(m_{1})=m_{1} * (k_{1} \oplus k_{2}) \oplus k_{0}^{+(L-1)}$, we can solve an unique $k_{1}$, thus the unique $k_{2}$ can be solved. \\
\indent
For other conditions (b)-(i), we can verify them in a similar way, we won't go into details here. \\
\indent
For the condition (i), As it will be used in the subsequent proof, we will describe it as a theorem here. \\
\\
\indent
\textbf{[Theorem 9]} As (4.5) defined, the following conclusions hold. \\
\indent
(i) For any equation $(*, m_{2}) \bigotimes (k_{1}, k_{2}) = (c_{1}, *)$, where $m_{2}$ is the unknown variable to be solved, $k_{1}, k_{2}, c_{1}$ are known variables, this equation can be seen as an identity, valid for all $m_{2}$. \\
\indent
(ii) For any equation $(*, *) \bigotimes (k_{1}, k_{2}) = (c_{1}, c_{2})$, where $c_{2}$ is the unknown variable to be solved, $k_{1}, k_{2}, c_{1}$ are known variables, this equation can be seen as an identity, valid for all $c_{2}$. \\
\\
\indent
If we construct an encryption algorithm where $(k_{1}, k_{2})$ is the key, $m_{1}$ is the plaintext, $c_{1}$ is the ciphertext, $m_{2}$ and $c_{2}$ never appear in the plaintext sequences or the ciphertext sequences, according to the condition (d), for a cracker, under the condition of only knowing the plaintext ciphertext correspondence $m_{1}, c_{1}$, any value $(k_{1}, k_{2})$ in the key space is valid and is indistinguishable, this feature allows us to construct the security that meets the requirements of computational complexity.  \\
\indent
We summarize it as the following theorem. \\
\\
\indent
\textbf{[Theorem 10]} As (4.5) defined, for any equation $(m_{1}, *) \bigotimes (k_{1}, k_{2}) = (c_{1}, *)$, which $(k_{1}, k_{2})$ are unknown variables and $m_{1}, c_{1}$ are known variables, this equation can be seen as an identity, valid for all $k_{1}, k_{2}$. \\
\indent
\begin{proof}
This theorem requires us to construct an algorithm that, given any $(k_{1}, k_{2})$ to find a valid equation $(m_{1}, *) \bigotimes (k_{1}, k_{2}) = (c_{1}, *)$. \\
\indent
When given any $(k_{1}, k_{2})$, we can solve $c_{2}$ and $m_{2}$ through the following method. \\
\begin{equation}
\notag
\begin{aligned}
c_{2}=\mathbb{E}_{k_{1}, k_{2}}(m_{1}) \\
m_{2}=(c_{1} \oplus c_{2}) / (k_{1} \oplus k_{2})
\end{aligned}
\end{equation}
\indent
Additionally, due to known $m_{1}, c_{1}$, each $m_{2}, c_{2}$ can solve an unique $k_{1}, k_{2}$, * in $(m_{1}, *) \bigotimes (k_{1}, k_{2}) = (c_{1}, *)$ can take all values without distinction, thus the equation $(m_{1}, *) \bigotimes (k_{1}, k_{2}) = (c_{1}, *)$ can be seen as an identity equation to unknown variables $k_{1}$ and $k_{2}$. \\
\end{proof}

\section{Construction of one-way functions}
\subsection{Introduction to one-way functions}
Before constructing the one-way function, we briefly introduce the properties of one-way function and the relationships with the P vs NP problem.\\
\\
\indent
\textbf{[Definition 1]} A function is a one-way function means that the function satisfies the following properties:\\
\indent
a) For a given  $x$, there exists a polynomial-time algorithm that output  $f(x)$.\\
\indent
b) Given $y$, it is difficult to find an  $x$   that satisfies  $y = f(x)$, that is, there does not exists a polynomial-time algorithm that finding the  $x$.\\
\indent
The  NP-complete  problem  refers  to  a  set  of  problems  that  are  verifiable   in polynomial-time  algorithm. For  all  NP-complete  problems,  whether  there  exists algorithms that are solvable in polynomial-time, this is the P vs NP problem. If $P \neq NP$,  then  for  some  NP  problems,  there  is  no  algorithm  that  is  solvable  in polynomial-time. If $P=NP$, then for all NP problems, there exists algorithms that are solvable in polynomial-time.\\
\indent
If one-way function exists, it means that there exists such an NP problem, which has no deterministic polynomial time solvable algorithm, that is, $P \neq NP$. This is a direct inference, which can be directly described as the following theorem, See \cite{REF9} for details.\\\newline
\indent
\textbf{[Theorem 11]} If one-way function exists, then $P \neq NP$.\\\newline

\indent
\subsection{Construction of one-way functions}
\indent
In this chapter, we constructed a short key symmetric encryption algorithm, which we named it Eagle. The key consists of 12 independent variables, each of which contains at least $2^{L-1}$ possible values. During the encryption process, several intermediate random numbers are generated at the encryption party, after the encryption process is completed, these intermediate random numbers are directly deleted at the encryption party, and only the ciphertext is sent to the decryption party. Under the condition of knowing the correspondence between any multiple sets of plaintext-ciphertext, the problem of cracking the key is equivalent to solving a system of equations about 12 variables together with some implicit variables. In solving the system of equations, we will prove that there is no universal algorithm without brute-force searching at least one variable, thus the computational complexity of cracking the key is at least $O(2^{L-1})$. \\
\indent
On the other hand, when the key known, both the encryption and decryption processes can be completed in linear time to the length of the key, meaning that the computational complexity for both encryption and decryption is $O(KL)$, where $K$ is a constant.\\
\indent
Furthermore, since the decryption process under the condition of knowing the ciphertext and key is reversible for the problem of cracking the key under the condition of knowing the correspondence between plaintext and ciphertext, the decryption process can be completed in $O(KL)$, while the complexity of  cracking the key is exponential, thus satisfying the condition of one-way function. \\
\indent
It should be noted that although the computational complexity of cracking the key is exponentially under the condition of knowing the correspondence between plaintext and ciphertext, this does not mean that this encryption algorithm is safe and practical. Assume the length of the key is $12L-12$ bits, although the computational complexity of cracking the key is not less than $O(2^{L-1})$, may be much lower than the expected complexity of $O(2^{12L-12})$. A secure and practical encryption algorithm needs to ensure that the cracking key can only exhaustively search for every possible key which is the issue to be considered in the next chapter. In this chapter, our goal is to design a one-way function where we only need to design and prove that the computational complexity for cracking keys is exponential to the length of the key. \\
\indent
Next, we will describe it in five sections. Section 5.2.1 introduces the key generation of the Eagle encryption algorithm, Section 5.2.2 introduces the process of the Eagle encryption algorithm, Section 5.2.3 introduces the process of the Eagle decryption algorithm, Section 5.2.4 analyzes the correctness and complexity of the Eagle encryption algorithm, and Section 5.2.5 provides the security analysis of the Eagle algorithm which give a construction of one-way function.\\

\indent
\subsubsection{$Eagle$ key generator}
\indent
The key is $(w_{1}, w_{2})$, $(w_{3}, w_{4})$, $(w_{5}, w_{6})$, $(w_{7}, w_{8})$, $(w_{9}, w_{10})$ and $(w_{11}, w_{12})$ where $w_{1}$, $w_{3}$, $w_{5}$, $w_{7}$, $w_{9}$, $w_{11}$ contains even number of bits that are 1, $w_{2}$, $w_{4}$, $w_{6}$, $w_{8}$, $w_{10}$, $w_{12}$ contains odd number of bits that are 1, thus $w_{1}$ and $w_{2}$ have odd number of bits different, $w_{3}$ and $w_{4}$ have odd number of bits different, $w_{5}$ and $w_{6}$ have odd number of bits different, $w_{7}$ and $w_{8}$ have odd number of bits different, $w_{9}$ and $w_{10}$ have odd number of bits different, $w_{11}$ and $w_{12}$ have odd number of bits different. \\
\indent
Assuming $w_{1}, w_{2}, w_{3}, w_{4}, w_{5}, w_{6}, w_{7}, w_{8}, w_{9}, w_{10}, w_{11}, w_{12}$ are all $L$ bits, the length of the entire key can be considered as $12L-12$ bits, this is because they have a loss of 12 bit. \\
\indent

\subsubsection{$Eagle$ encryption process}
\indent
The plaintext $P$ is grouped by $L-1$ bits, and the last group with less than $L-1$ bits are randomly filled into $L-1$ bits, then fill one bit with 1 to the last bit of each group to ensure the length of each group is $L$ bits. The total number of groups is assumed to be $K$, denoted as $P=(P_{1}, P_{2}, ..., P_{K})$. Obviously, $P_{i}$ are all odd numbers. \\
\indent
Then generate random numbers of $18K+3$ groups with each have $L$ bits, denoted as $(t_{1}$, $t_{2}$, ..., $t_{6K}$, $t_{6K+1})$, $(x_{1}$, $x_{2}$, ..., $x_{6K}$, $x_{6K+1})$, $(y_{1}$, $y_{2}$, ..., $y_{6K}$, $y_{6K+1})$. where $x_{i}(1 \le i \le 6K+1)$, $y_{i}(1 \le i \le 6K+1)$, $t_{i}(1 \le i \le 6K+1)$ are all odd numbers. \\
\indent
Generate another five random numbers with each have $L$ bits denoted as $\mu_{1}$, $u_{1}$, $v_{1}$, $u_{2}$ and $v_{2}$, where $\mu_{1}$, $u_{1}$, $v_{1}$, $u_{2}$ and $v_{2}$ are all odd numbers. \\
\indent
Next, we need to define two simple functions $\phi(x), \psi(x)$ as follows.\\
\begin{algorithm} [H]
\caption{calculation of $\phi(x)$.}
\label{alg: Framework}
\begin{algorithmic}[1]
\REQUIRE ~~ \\
Input: $x$. \\
\ENSURE ~~ \\
Output: $\phi(x)$. \\
\STATE $y \leftarrow 0$. \\
\STATE $y[1] \leftarrow x[1]$ \\
\FOR{$i=2; i \leq L; i++$}
	\STATE $y[i]=x[1] \oplus x[2] \oplus ... \oplus x[i]$ \\
\ENDFOR
\RETURN $y$. \\
\end{algorithmic}
\end{algorithm}
\indent
Take the first bit of $x$ as the first bit of $\phi(x)$, take the XOR of the first two bits of $x$ as the second bit of $\phi(x)$, take the XOR of the first three bits of $x$ as the third bit of $\phi(x)$, and so on, to obtain each bit of $\phi(x)$. \\
\indent
It is not difficult to find that $\phi(x)$ changes $x$ to an odd number if $x$ contains odd number of bits that are 1, and $\phi(x)$ changes $x$ to an even number if $x$ contains even number of bits that are 1.\\

\begin{algorithm} [H]
\caption{calculation of $\psi(x)$.}
\label{alg: Framework}
\begin{algorithmic}[1]
\REQUIRE ~~ \\
Input: $x$. \\
\ENSURE ~~ \\
Output: $\psi(x)$. \\
\STATE $y \leftarrow 0$. \\
\STATE $y[1] \leftarrow x[1]$. \\
\FOR{$i = 2; i \leq L; i++$}
	\STATE $y[i] \leftarrow x[i-1] \oplus x[i]$ \\
\ENDFOR
\RETURN $y$. \\
\end{algorithmic}
\end{algorithm}
\indent
$\psi(x)$ is the inverse function of $\phi(x)$, take the first bit of $\phi(x)$ as the first bit of $x$, take the XOR of the first bit and the second bit of $\phi(x)$ as the second bit of $x$, take the XOR of the second bit and the third bit of $\phi(x)$ as the third bit of $x$, and so on, to obtain each bit of $x$ from $\phi(x)$. \\  
\indent
It is not difficult to find that $\psi(x)$ changes $x$ to a number containing odd number of bits that are 1 if $x$ is an odd number, and $\psi(x)$ changes $x$ to a number containing even number of bits that are 1 if $x$ is an even number.\\
\indent
$\phi(x)$ and $\psi(x)$ are inverse functions of each other. \\
\indent
For convenience, we introduce another function $\theta(x)$, which defined as $\theta(x)=x^{+(L-1)}$, do left cycle shift of $x$ by $L-1$ bits. \\
\indent
It is not difficult to verify that $\phi(x), \psi(x), \theta(x)$ are all linear functions, that is $\phi(x_{1} \oplus x_{2})=\phi(x_{1}) \oplus \phi(x_{2})$, $\psi(x_{1} \oplus x_{2})=\psi(x_{1}) \oplus \psi(x_{2})$, $\theta(x_{1} \oplus x_{2})=\theta(x_{1}) \oplus \theta(x_{2})$ hold for any $x_{1}$ and $x_{2}$. \\
\indent
Additionally, we introduce an operator $\oplus 1$ to change the parity of a number, or changes a number to containing an even number of bits that are 1 if the number contains an odd number of bits that are 1, or changes a number to containing an odd number of bits that are 1 if the number contains an even number of bits that are 1. \\
\\
\indent
The encryption process is described in the following algorithm. \\
\begin{breakablealgorithm} 
\caption{$Eagle$-Encrypt.}
\label{alg: Framework}
\begin{algorithmic}[1]
\REQUIRE ~~ \\
Key: $(w_{1}, w_{2})$, $(w_{3}, w_{4})$, $(w_{5}, w_{6})$, $(w_{7}, w_{8})$, $(w_{9}, w_{10})$, $(w_{11}, w_{12})$. \\
Plaintext sequence: $P=(P_{1}, P_{2}, ..., P_{K})$. \\
Random number sequence: \\
\,\,\,\,\,\,\,\,\,\,\,\,\,\,\,\,$(t_{1}, t_{2}, ..., t_{6K+1})$, \\
\,\,\,\,\,\,\,\,\,\,\,\,\,\,\,\,$(x_{1}, x_{2}, ..., x_{6K+1})$, \\
\,\,\,\,\,\,\,\,\,\,\,\,\,\,\,\,$(y_{1}, y_{2}, ..., y_{6K+1})$. \\
Random number: $\mu_{1}, u_{1}, v_{1}, u_{2}, v_{2}$. \\
\ENSURE ~~ \\
Ciphertext sequence:  \\
\,\,\,\,\,\,\,\,\,\,\,\,\,\,\,\,$(C_{1}, C_{2}, ..., C_{6K+1})$, \\
\,\,\,\,\,\,\,\,\,\,\,\,\,\,\,\,$(x_{1}, x_{2}, ..., x_{6K+1})$, \\
\,\,\,\,\,\,\,\,\,\,\,\,\,\,\,\,$(y_{1}, y_{2}, ..., y_{6K+1})$, \\
\,\,\,\,\,\,\,\,\,\,\,\,\,\,\,\,$(z_{1}, z_{2}, ..., z_{K})$, \\
\,\,\,\,\,\,\,\,\,\,\,\,\,\,\,\,$u_{1}, v_{1}, u_{2}, v_{2}, t_{6K+1}$. \\
\STATE $w_{u_{1}} \leftarrow \mathbb{E}_{w_{9}, w_{10}}(u_{1}) \oplus 1$ \\
\STATE $w_{v_{1}} \leftarrow \mathbb{E}_{w_{9}, w_{10}}(v_{1})$ \\
\STATE $w_{u_{2}} \leftarrow \mathbb{E}_{w_{11}, w_{12}}(u_{2}) \oplus 1$ \\
\STATE $w_{v_{2}} \leftarrow \mathbb{E}_{w_{11}, w_{12}}(v_{2})$ \\
\FOR{$j=1; j \le 6K+1; j++$}
	\IF{$j \equiv 1 \mod 6$}
	\STATE $w_{x_{j}} \leftarrow \mathbb{E}_{w_{1}, w_{2}}(x_{j}) \oplus 1$ \\
	\STATE $w_{y_{j}} \leftarrow \mathbb{E}_{w_{3}, w_{4}}(y_{j})$ \\
	\ELSE
	\STATE $w_{x_{j}} \leftarrow \mathbb{E}_{w_{5}, w_{6}}(x_{j}) \oplus 1$ \\
	\STATE $w_{y_{j}} \leftarrow \mathbb{E}_{w_{7}, w_{8}}(y_{j})$ \\
	\ENDIF
\ENDFOR
\FOR{$i=1; i \le K; i++$}
	\STATE $z_{i} \leftarrow P_{i} \oplus t_{6i - 5}$ \\
	\STATE $(C_{6i-5}, \lambda_{6i-5}) \leftarrow (t_{6i-5}, \mu_{6i-5}) \bigotimes (w_{x_{6i-5}}, w_{y_{6i-5}})$ \\
	\STATE $\lambda_{6i-5} \leftarrow \phi(\lambda_{6i-5})$ \\
	\STATE $\mu_{6i-4} \leftarrow \mathbb{E}_{w_{u_{1}}, w_{v_{1}}}(\lambda_{6i-5})$
	\STATE $\mu_{6i-4} \leftarrow \phi(\mu_{6i-4})$ \\
	\STATE $(C_{6i-4}, \lambda_{6i-4}) \leftarrow (t_{6i-4}, \mu_{6i-4}) \bigotimes (w_{x_{6i-4}}, w_{y_{6i-4}})$ \\
	\STATE $\lambda_{6i-4} \leftarrow \phi(\lambda_{6i-4})$ \\
	\STATE $\mu_{6i-3} \leftarrow \mathbb{E}_{w_{u_{2}}, w_{v_{2}}}(\lambda_{6i-4})$
	\STATE $\mu_{6i-3} \leftarrow \phi(\mu_{6i-3})$ \\	
	\STATE $(C_{6i-3}, \lambda_{6i-3}) \leftarrow (t_{6i-3}, \mu_{6i-3}) \bigotimes (w_{x_{6i-3}}, w_{y_{6i-3}})$ \\
	\STATE $\lambda_{6i-3} \leftarrow \phi(\lambda_{6i-3})$ \\
	\STATE $\mu_{6i-2} \leftarrow \mathbb{E}_{w_{u_{2}}, w_{v_{2}}}(\lambda_{6i-3})$
	\STATE $\mu_{6i-2} \leftarrow \phi(\mu_{6i-2})$ \\	
	\STATE $(C_{6i-2}, \lambda_{6i-2}) \leftarrow (t_{6i-2}, \mu_{6i-2}) \bigotimes (w_{x_{6i-2}}, w_{y_{6i-2}})$ \\
	\STATE $\lambda_{6i-2} \leftarrow \phi(\lambda_{6i-2})$ \\
	\STATE $\mu_{6i-1} \leftarrow \mathbb{E}_{w_{u_{2}}, w_{v_{2}}}(\lambda_{6i-2})$
	\STATE $\mu_{6i-1} \leftarrow \phi(\mu_{6i-1})$ \\	
	\STATE $(C_{6i-1}, \lambda_{6i-1}) \leftarrow (t_{6i-1}, \mu_{6i-1}) \bigotimes (w_{x_{6i-1}}, w_{y_{6i-1}})$ \\
	\STATE $\lambda_{6i-1} \leftarrow \phi(\lambda_{6i-1})$ \\
	\STATE $\mu_{6i} \leftarrow \mathbb{E}_{w_{u_{2}}, w_{v_{2}}}(\lambda_{6i-1})$
	\STATE $\mu_{6i} \leftarrow \phi(\mu_{6i})$ \\	
	\STATE $(C_{6i}, \lambda_{6i}) \leftarrow (t_{6i}, \mu_{6i}) \bigotimes (w_{x_{6i}}, w_{y_{6i}})$ \\
	\STATE $\lambda_{6i} \leftarrow \phi(\lambda_{6i})$ \\
	\STATE $\mu_{6i+1} \leftarrow \mathbb{E}_{w_{u_{2}}, w_{v_{2}}}(\lambda_{6i})$
	\STATE $\mu_{6i+1} \leftarrow \phi(\mu_{6i+1})$ \\
\ENDFOR
\STATE $(C_{6K+1}, \lambda_{6K+1}) \leftarrow (t_{6K+1}, \mu_{6K+1}) \bigotimes (w_{x_{6K+1}}, w_{y_{6K+1}})$ \\
\end{algorithmic}
\end{breakablealgorithm}
\indent
The above algorithm can be represented by the following figure. \\
\begin{tikzcd}
\mu_{1} \arrow[dd, "t_{1} => z_{1}=P_{1}\oplus t_{1}"]                                      &                                                                           &    &                                                                  \\
                                                                                            & {} \arrow[r, "{x_{1}\rightarrow w_{x_{1}},y_{1}\rightarrow w_{y_{1}}}"]   & {} & {(t_{1}, \mu_{1}) \bigotimes (w_{x_{1}}, w_{y_{1}})} \arrow[lld] \\
{\mu_{2}=\mathbb{E}_{w_{u_{1}},w_{v_{1}}}(\lambda_{1})} \arrow[dd, "t_{2}"]                 & {(C_{1},\lambda_{1})} \arrow[l]                                           &    &                                                                  \\
                                                                                            & {} \arrow[r, "{x_{2}\rightarrow w_{x_{2}},y_{2}\rightarrow w_{y_{2}}}"]   & {} & {(t_{2}, \mu_{2}) \bigotimes (w_{x_{2}}, w_{y_{2}})} \arrow[lld] \\
{\mu_{3}=\mathbb{E}_{w_{u_{2}},w_{v_{2}}}(\lambda_{2})} \arrow[dd, "t_{3}"]                 & {(C_{2},\lambda_{2})} \arrow[l]                                           &    &                                                                  \\
                                                                                            & {} \arrow[r, "{x_{3}\rightarrow w_{x_{3}},y_{3}\rightarrow w_{y_{3}}}"]   & {} & {(t_{3}, \mu_{3}) \bigotimes (w_{x_{3}}, w_{y_{3}})} \arrow[lld] \\
{\mu_{4}=\mathbb{E}_{w_{u},w_{v}}(\lambda_{3})} \arrow[dd, "t_{4}"]                         & {(C_{3},\lambda_{3})} \arrow[l]                                           &    &                                                                  \\
                                                                                            & {} \arrow[r, "{x_{4}\rightarrow w_{x_{4}},y_{4}\rightarrow w_{y_{4}}}"]   & {} & {(t_{4}, \mu_{4}) \bigotimes (w_{x_{4}}, w_{y_{4}})} \arrow[lld] \\
{\mu_{5}=\mathbb{E}_{w_{u_{2}},w_{v_{2}}}(\lambda_{4})} \arrow[dd, "t_{5}"]                 & {(C_{4},\lambda_{4})} \arrow[l]                                           &    &                                                                  \\
                                                                                            & {} \arrow[r, "{x_{5}\rightarrow w_{x_{5}},y_{5}\rightarrow w_{y_{5}}}"]   & {} & {(t_{5},\mu_{5})\bigotimes(w_{x_{5}},w_{y_{5}})} \arrow[lld]     \\
{\mu_{6}=\mathbb{E}_{w_{u_{2}},w_{v_{2}}}(\lambda_{5})} \arrow[dd, "t_{6}"]                 & {(C_{5},\lambda_{5})}                                                     &    &                                                                  \\
                                                                                            & {} \arrow[r, "{x_{6}\rightarrow w_{x_{6}}, y_{6} \rightarrow w_{y_{6}}}"] & {} & {(t_{6},\mu_{6}) \bigotimes (w_{x_{6}}, w_{y_{6}})} \arrow[lld]  \\
{\mu_{7}=E_{w_{u_{2}},w_{v_{2}}}(\lambda_{6})} \arrow[dd, "t_{2}=>z_{2}=P_{2}\oplus t_{7}"] & {(C_{6},\lambda_{6})}                                                     &    &                                                                  \\
                                                                                            & {} \arrow[r, "{x_{7}\rightarrow w_{x_{7}},y_{7}\rightarrow w_{y_{7}}}"]   & {} & {(t_{7},\mu_{7})\bigotimes(w_{x_{7}},w_{y_{7}})} \arrow[lld]     \\
{\mu_{8}=\mathbb{E}_{w_{u_{1}},w_{v_{1}}}(\lambda_{7})}                                     & {(C_{7},\lambda_{7})}                                                     &    &                                                                 
\end{tikzcd}
\\
\\
\indent
It is not difficult to find that $C_{i}$ all contain even number of bits that are 1. $z_{i}$ are all even numbers. \\
\indent
Additionally, it should be noted that all intermediate variables as $\lambda_{i}$ and $\mu_{i}$ do not appear in the ciphertext, but are directly deleted by the encrypting party after the encryption algorithm is executed. \\

\subsubsection{$Eagle$ decryption process}
\indent
For the decrypting party, the key is also $(w_{1}, w_{2})$, $(w_{3}, w_{4})$, $(w_{5}, w_{6})$, $(w_{7}, w_{8})$, $(w_{9}, w_{10})$, $(w_{11}, w_{12})$, the ciphertext is $(C_{1}$, $C_{2}$, ..., $C_{6K+1})$, $(x_{1}$, $x_{2}$,..., $x_{6K+1})$, $(y_{1}$, $y_{2}$, ..., $y_{6K+1})$, $(z_{1}$, $z_{2}$, ..., $z_{K})$, $u_{1}, v_{1}, u_{2}, v_{2}, t_{6K+1}$.  Please refer to the following for details. \\
\begin{breakablealgorithm}
\caption{$Eagle$-Decrypt.}
\label{alg: Framework}
\begin{algorithmic}[1]
\REQUIRE ~~ \\
Key: $(w_{1}, w_{2}), (w_{3}, w_{4}), (w_{5}, w_{6}), (w_{7}, w_{8}), (w_{9}, w_{10}), (w_{11}, w_{12})$. \\
Ciphertext sequence: \\
\,\,\,\,\,\,\,\,\,\,\,\,\,\,\,\,$(C_{1}, C_{2}, ..., C_{6K+1})$, \\
\,\,\,\,\,\,\,\,\,\,\,\,\,\,\,\,$(x_{1}, x_{2}, ..., x_{6K+1})$, \\
\,\,\,\,\,\,\,\,\,\,\,\,\,\,\,\,$(y_{1}, y_{2}, ..., y_{6K+1})$, \\
\,\,\,\,\,\,\,\,\,\,\,\,\,\,\,\,$(z_{1}, z_{2}, ..., z_{K})$, \\
\,\,\,\,\,\,\,\,\,\,\,\,\,\,\,\,$u_{1}, v_{1}, u_{2}, v_{2}, t_{6K+1}$. \\
\ENSURE ~~ \\
Plaintext sequence: $P=(P_{1}, P_{2}, ..., P_{K})$.
\STATE $w_{u_{1}} \leftarrow \mathbb{E}_{w_{9}, w_{10}}(u_{1}) \oplus 1$ \\
\STATE $w_{v_{1}} \leftarrow \mathbb{E}_{w_{9}, w_{10}}(v_{1})$ \\
\STATE $w_{u_{2}} \leftarrow \mathbb{E}_{w_{11}, w_{12}}(u_{2}) \oplus 1$ \\
\STATE $w_{v_{2}} \leftarrow \mathbb{E}_{w_{11}, w_{12}}(v_{2})$ \\
\FOR{$j=1; j \le 6K+1; j++$}
	\IF{$j \equiv 1 \mod 6$}
	\STATE $w_{x_{j}} \leftarrow \mathbb{E}_{w_{1}, w_{2}}(x_{j}) \oplus 1$ \\
	\STATE $w_{y_{j}} \leftarrow \mathbb{E}_{w_{3}, w_{4}}(y_{j})$ \\
	\ELSE
	\STATE $w_{x_{j}} \leftarrow \mathbb{E}_{w_{5}, w_{6}}(x_{j}) \oplus 1$ \\
	\STATE $w_{y_{j}} \leftarrow \mathbb{E}_{w_{7}, w_{8}}(y_{j})$ \\
	\ENDIF
\ENDFOR
\STATE $\lambda_{6K+1} \leftarrow \mathbb{E}_{w_{x_{6K+1}}, w_{y_{6K+1}}}(t_{6K+1})$ \\
\STATE $\mu_{6K+1} \leftarrow (\lambda_{6K+1} \oplus C_{6K+1}) / (w_{x_{6K+1}} \oplus w_{y_{6K+1}})$ \\
\STATE $\mu_{6K+1} \leftarrow \psi(\mu_{6K+1})$ \\
\STATE $\lambda_{6K} \leftarrow \mathbb{D}_{w_{u_{2}},w_{v_{2}}}(\mu_{6K+1})$ \\
\STATE $\lambda_{6K} \leftarrow \psi(\lambda_{6K})$ \\
\FOR{$i=K; i \ge 1; i--$}
	\STATE $\mu_{6i} \leftarrow (\lambda_{6i} \oplus C_{6i}) / (w_{x_{6i}} \oplus w_{y_{6i}})$ \\
	\STATE $\mu_{6i} \leftarrow \psi(\mu_{6i})$ \\
	\STATE $\lambda_{6i-1} \leftarrow \mathbb{D}_{w_{u_{2}}, w_{v_{2}}}(\mu_{6i})$ \\
	\STATE $\lambda_{6i-1} \leftarrow \psi(\lambda_{6i-1})$ \\
	\STATE $\mu_{6i-1} \leftarrow (\lambda_{6i-1} \oplus C_{6i-1}) / (w_{x_{6i-1}} \oplus w_{y_{6i-1}})$ \\
	\STATE $\mu_{6i-1} \leftarrow \psi(\mu_{6i-1})$ \\
	\STATE $\lambda_{6i-2} \leftarrow \mathbb{D}_{w_{u_{2}}, w_{v_{2}}}(\mu_{6i-1})$ \\
	\STATE $\lambda_{6i-2} \leftarrow \psi(\lambda_{6i-2})$ \\
	\STATE $\mu_{6i-2} \leftarrow (\lambda_{6i-2} \oplus C_{6i-2}) / (w_{x_{6i-2}} \oplus w_{y_{6i-2}})$ \\
	\STATE $\mu_{6i-2} \leftarrow \psi(\mu_{6i-2})$ \\
	\STATE $\lambda_{6i-3} \leftarrow \mathbb{D}_{w_{u_{2}}, w_{v_{2}}}(\mu_{6i-2})$ \\
	\STATE $\lambda_{6i-3} \leftarrow \psi(\lambda_{6i-3})$ \\	
	\STATE $\mu_{6i-3} \leftarrow (\lambda_{6i-3} \oplus C_{6i-3}) / (w_{x_{6i-3}} \oplus w_{y_{6i-3}})$ \\
	\STATE $\mu_{6i-3} \leftarrow \psi(\mu_{6i-3})$ \\
	\STATE $\lambda_{6i-4} \leftarrow \mathbb{D}_{w_{u_{2}}, w_{v_{2}}}(\mu_{6i-3})$ \\
	\STATE $\lambda_{6i-4} \leftarrow \psi(\lambda_{6i-4})$ \\
	\STATE $\mu_{6i-4} \leftarrow (\lambda_{6i-4} \oplus C_{6i-4}) / (w_{x_{6i-4}} \oplus w_{y_{6i-4}})$ \\
	\STATE $\mu_{6i-4} \leftarrow \psi(\mu_{6i-4})$ \\
	\STATE $\lambda_{6i-5} \leftarrow \mathbb{D}_{w_{u_{1}}, w_{v_{1}}}(\mu_{6i-4})$ \\
	\STATE $\lambda_{6i-5} \leftarrow \psi(\lambda_{6i-5})$ \\
	\STATE $t_{6i-5} \leftarrow \mathbb{D}_{w_{x_{6i-5}}, w_{y_{6i-5}}}(\lambda_{6i-5})$ \\
	\STATE $P_{i} \leftarrow z_{i} \oplus t_{6i-5}$ \\
	\STATE $\mu_{6i-5} \leftarrow (\lambda_{6i-5} \oplus C_{6i-5}) / (w_{x_{6i-5}} \oplus w_{y_{6i-5}})$ \\
	\STATE $\mu_{6i-5} \leftarrow \psi(\mu_{6i-5})$ \\
	\STATE $\lambda_{6i-6} \leftarrow \mathbb{D}_{w_{u_{2}}, w_{v_{2}}}(\mu_{6i-5})$ \\
	\STATE $\lambda_{6i-6} \leftarrow \psi(\lambda_{6i-6})$ \\
\ENDFOR
\end{algorithmic}
\end{breakablealgorithm}

\indent
\subsubsection{Basic analysis of $Eagle$}
\indent
It is obvious that $Eagle$ encryption algorithm is correct because the decryption process and encryption process are mutually inverse.  \\
\indent
From the encryption process above, it can be seen that the $Eagle$ encryption algorithm is divided into two stages: the preparation stage and the encryption stage. In the preparation stage, different parameters  are randomly generated. In the encryption stage, the ciphertexts generated by the random parameters and the plaintexts together with the keys. After the encryption process is completed, some intermediate random parameters are directly deleted by the encryption party and do not appear in the ciphertexts. \\
\indent
It should also be noted that after being encrypted by the $Eagle$ encryption algorithm, the length of the ciphertext is about 19 times the length of the plaintext. This is because in the preparation stage of the $Eagle$ encryption algorithm, independent random numbers with a length of about 18 times the plaintext are generated, and then some random numbers are directly packaged into the ciphertexts, and some random numbers binded into the final ciphertexts. \\
\indent
The $Eagle$ encryption algorithm performs bit by bit in encryption and decryption process, and the computational complexity of operation for each bit is also a constant level. Therefore, the encryption and decryption complexity of the $Eagle$ encryption algorithm can be regarded as $O(KL)$. \\

\subsubsection{A construction of one-way function}
\indent
When given the ciphertext ($C_{1}$, $C_{2}$, ..., $C_{6K+1}$), ($x_{1}$, $x_{2}$, ..., $x_{6K+1}$), ($y_{1}$, $y_{2}$, ..., $y_{6K+1}$), ($z_{1}$, $z_{2}$, ..., $z_{K}$),  and $(u_{1}, v_{1}, u_{2}, v_{2}, t_{6K+1})$ of the $Eagle$ encryption algorithm, we define the following function. \\
\begin{equation}
f(w_{1}, w_{2}, w_{3}, w_{4}, w_{5}, w_{6}, w_{7}, w_{8}, w_{9}, w_{10}, w_{11}, w_{12}) = (P_{1}, P_{2}, ..., P_{K}) \tag{5.1}
\end{equation}
\indent
where $(w_{1}, w_{2}, w_{3}, w_{4}, w_{5}, w_{6}, w_{7}, w_{8}, w_{9}, w_{10}, w_{11}, w_{12})$ is the short key and $(P_{1}$, $P_{2}$, ..., $P_{K})$ is the plaintext decrypted by the $Eagle$ decryption algorithm. \\
\indent
The inverse function of $f$ can be defined as \\
\begin{equation}
\tag{5.2}
\begin{aligned}
f^{-1}(P_{1}, P_{2}, ..., P_{K})=\{(w_{1}, w_{2}, w_{3}, w_{4}, w_{5}, w_{6}, w_{7}, w_{8}, w_{9}, w_{10}, w_{11}, w_{12}) | \\
 f(w_{1}, w_{2}, w_{3}, w_{4}, w_{5}, w_{6}, w_{7}, w_{8}, w_{9}, w_{10}, w_{11}, w_{12}) = (P_{1}, P_{2}, ..., P_{K})\}   
\end{aligned}
\end{equation}
\indent
Solving $f^{-1}$ is equivalent to finding the key $(w_{1}$, $w_{2}$, $w_{3}$, $w_{4}$, $w_{5}$, $w_{6}$, $w_{7}$, $w_{8}$, $w_{9}$, $w_{10}$, $w_{11}$, $w_{12})$ to match the plaintext $(P_{1}, P_{2}, ..., P_{K})$. \\
\indent
In fact, if we can prove that we can't solve $f^{-1}$ in polynomial time, that is, for any attacker who only knows the plaintext-ciphertext pairs, there is no effective method in polynomial time to find the correct key. \\
\indent
Here we need to clarify that we only need to prove that we can't solve $w=$$(w_{1}$, $w_{2}$, $w_{3}$, $w_{4}$, $w_{5}$, $w_{6}$, $w_{7}$, $w_{8}$, $w_{9}$, $w_{10}$, $w_{11}$, $w_{12})$ in polynomial time, instead of relying only exhaustive search to find $w$. In fact,``only exhaustive search can be used to find $w$" is much stricter semantically than “unable to solve $w$ in polynomial time”. In this chapter, we only need to prove that $w$ can't be solved in polynomial time, and this already satisfies the condition of one-way function. \\
\indent
Intuitively, we can't solve $(w_{1}, w_{2})$ when only $(x_{1}$, $x_{7}$, $x_{13}$, ..., $x_{6K+1})$ are given, we can't solve $(w_{3}, w_{4})$ when only $(y_{1}$, $y_{7}$, $y_{13}$, ..., $y_{6K+1})$ are given, we can't solve $(w_{5}, w_{6})$ when only $(x_{2}$, $x_{3}$, $x_{4}$, $x_{5}$, $x_{6}$, $x_{8}$, ..., $x_{6K})$ are given, we can't solve $(w_{7}, w_{8})$ when only $(y_{2}$, $y_{3}$, $y_{4}$, $y_{5}$, $y_{6}$, $y_{8}$, ..., $y_{6K})$ are given, we can't solve $(w_{9}, w_{10})$ when only $u_{1}, v_{1}$ are given, we can't solve $w_{11}, w_{12}$ when only $u_{2}, v_{2}$ are given, when given$P_{i}, z_{i}, C_{i}$, we can obtain the equation $(P_{i} \oplus z_{i}, *) \bigotimes (w_{x_{i}}, w_{y_{i}}) = (C_{i}, *)$, according to the theorem 10, this is an identity equation. \\
\indent
Based on the above analysis, given the correspondence between the plaintext ($P_{1}, P_{2}, ..., P_{K}$) and ciphertext ($C_{1}$, $C_{2}$, ..., $C_{6K+1}$), $(x_{1}$, $x_{2}$,..., $x_{6K+1})$, $(y_{1}$, $y_{2}$, ..., $y_{6K+1})$, $(z_{1}$, $z_{2}$,..., $z_{K})$, $u_{1}, v_{1}, u_{2}, v_{2}, t_{6K+1}$ of the Eagle encryption algorithm, the problem of solving the key $(w_{1} $, $w_{2}$, $w_{3}$, $w_{4}$, $w_{5}$, $w_{6}$, $w_{7}$, $w_{8}$, $w_{9}$, $w_{10}$, $w_{11}$, $w_{12})$ is equivalent to the problem of solving the respective keys in six independent encryption algorithms under the condition that only their plaintext is known. The first encryption algorithm solves its key $(w_{9}, w_{10})$ under the condition that its plaintext $(u_{1}, v_{1})$ are known. For the second encryption algorithm, the key $(w_{11}, w_{12})$ need to be solved under the condition that only the plaintext $(u_{2}, v_{2})$ is known. For the third encryption algorithm, the key $(w_{1}, w_{2})$ need to be solved under the condition that only the plaintext ($x_{1}$, $x_{7}$, ..., $x_{6K+1}$) is known. For the forth encryption algorithm, the key $(w_{3}, w_{4})$ need to be solved under the condition that only the plaintext ($y_{1}$, $y_{7}$, ..., $y_{6K+1}$) is known, etc. \\
\indent
In fact, for these six independent encryption algorithms, under the condition of only knowing their respective plaintext, every possible key cannot be reasonably excluded, and every possible key must be exhaustively searched. From the perspective of the entire Eagle encryption algorithm, at least one set of keys from $(w_{1}$, $w_{2})$ or $(w_{3}$, $w_{4})$ or $(w_{5}, w_{6})$ or $(w_{7}, w_{8})$ or $(w_{9}, w_{10})$ or $(w_{11}, w_{12})$ needs to be exhaustively searched to find a valid key. The computational complexity of exhausting either $(w_{1}$, $w_{2})$ or $(w_{3}$, $w_{4})$ or $(w_{5}, w_{6})$ or $(w_{7}, w_{8})$ or $(w_{9}, w_{10})$ or $(w_{11}, w_{12})$ is exponentially.  \\
\indent
During the decryption process, all intermediate variables as $\lambda_{i}, \mu_{i}$ are unknown in the ciphertext, and these intermediate variables need to be gradually restored from the last group to the first group. \\
\indent
Without specifying any intermediate variables such as  $\lambda_{i}$ or $\mu_{i}$, the problem of solving $f^{-1}$ is equivalently transformed into solving a system of equations about the unknown variables as $(w_{1}$, $w_{2}$, $w_{3}$, $w_{4}$, $w_{5}$, $w_{6}$, $w_{7}$. $w_{8}$, $w_{9}$, $w_{10}$, $w_{11}$, $w_{12})$, any equation in this system of equations can be regarded as an identity equation about the unknown variables, and it is impossible to find an effective solution.\\
\indent
On the other hand, given $(P_{1}$, $P_{2}$, ... , $P_{K})$ and ($C_{1}$, $C_{2}$, ..., $C_{6K+1}$), $(x_{1}$, $x_{2}$,..., $x_{6K+1})$, $(y_{1}$, $y_{2}$, ..., $y_{6K+1})$, $(z_{1}$, $z_{2}$, ..., $z_{K})$, $u_{1}, v_{1}, u_{2}, v_{2}, t_{6K+1}$, every intermediate variable $\lambda_{i}$ or $\mu_{i}$ can take any value, the computational complexity of traversing every intermediate variable is $O(2^{L-1})$. \\
\indent
This indicates that the computational complexity of solving $f^{-1}$ is at least $O(2^{L-1})$. \\
\indent
Let's give the following theorem. \\

\indent
\textbf{[Theorem 12]} When given the ciphertext ($C_{1}$, $C_{2}$, ..., $C_{6K+1}$), $(x_{1}$, $x_{2}$,..., $x_{6K+1})$, $(y_{1}$, $y_{2}$, ..., $y_{6K+1})$, $(z_{1}$, $z_{2}$, ..., $z_{K})$, $u_{1}, v_{1}, u_{2}, v_{2}, t_{6K+1}$ and the plaintext $(P_{1}$, $P_{2}$, ..., $P_{K})$ of the $Eagle$ encryption algorithm, the computational complexity of the problem of cracking the key $(w_{1}$, $w_{2}$, $w_{3}$, $w_{4}$, $w_{5}$, $w_{6}$, $w_{7}$, $w_{8}$, $w_{9}$, $w_{10}$, $w_{11}$, $w_{12})$ is at least $O(2^{L-1})$. \\
\indent
\begin{proof}
\indent
Firstly, the problem of cracking keys $(w_{1}$, $w_{2}$, $w_{3}$, $w_{4}$, $w_{5}$, $w_{6}$, $w_{7}$, $w_{8}$, $w_{9}$, $w_{10}$, $w_{11}$, $w_{12})$ can be equivalently transformed into solving a system of equations. \\
\indent
By observing the structure of Algorithm 7 and Algorithm 8, when $i\equiv 1 \pmod{6}$, the plaintext ciphertext pair of the $i$-th group satisfy the equation $(P_{i} \oplus z_{i}, \mu_{i}) \bigotimes (w_{x_{i}}, w_{y_{i}})=(C_{i}, \lambda_{i})$, where $\mu_{i}$, $\lambda_{i}$ are used to represent the intermediate variables generated by the encryption party and $w_{x_{i}}=\mathbb{E}_{w_{1}, w_{2}}(x_{i})$ and $w_{y_{i}}=\mathbb{E}_{w_{3}, w_{4}}(y_{i})$. When $i \not \equiv 1 \pmod{6}$, the plaintext ciphertext pair of the $i$-th group satisfy the equation $(*, \mu_{i}) \bigotimes (w_{x_{i}}, w_{y_{i}})=(C_{i}, \lambda_{i})$ and $w_{x_{i}}=\mathbb{E}_{w_{5}, w_{6}}(x_{i})$ and $w_{y_{i}}=\mathbb{E}_{w_{7}, w_{8}}(y_{i})$. The structure of the entire parameter can be represented as shown in the figure below. \\
\begin{tikzcd}
P_{1}\oplus z_{1}                                          & \mu_{1}                                                                                                      &  & *                                                          & \mu_{2}                                                                                         &  & *                                                          & \mu_{3}     \\
C_{1}                                                      & \lambda_{1} \arrow[rrru, "{\mu_{2}=\phi(\mathbb{E}_{w_{u_{1}},w_{v_{1}}}(\phi(\lambda_{1})))}", shift right] &  & C_{2}                                                      & \lambda_{2} \arrow[rrru, "{\mu_{3}=\phi(\mathbb{E}_{w_{u_{2}},w_{v_{2}}}(\phi(\lambda_{2})))}"] &  & C_{3}                                                      & \lambda_{3} \\
{} \arrow[r, "{w_{x_{1}},w_{y_{1}}}", no head, shift left] & {}                                                                                                           &  & {} \arrow[r, "{w_{x_{2}},w_{y_{2}}}", no head, shift left] & {}                                                                                              &  & {} \arrow[r, "{w_{x_{3}},w_{y_{3}}}", no head, shift left] & {}         
\end{tikzcd}
\\
\indent
Where $w_{x_{i}}=\mathbb{E}_{w_{1}, w_{2}}(x_{i}) \oplus 1$ and $w_{y_{i}}=\mathbb{E}_{w_{3}, w_{4}}(y_{i})$ for $i \equiv 1 \mod 6$.  $w_{x_{i}}=\mathbb{E}_{w_{5}, w_{6}}(x_{i}) \oplus 1$ and $w_{y_{i}}=\mathbb{E}_{w_{7}, w_{8}}(y_{i})$ for $i \not \equiv 1 \mod 6$. $w_{u_{1}}=\mathbb{E}_{w_{9}, w_{10}}(u_{1})$, $w_{v_{1}}=\mathbb{E}_{w_{9}, w_{10}}(v_{1})$, $w_{u_{2}}=\mathbb{E}_{w_{11}, w_{12}}(u_{2})$, $w_{v_{2}}=\mathbb{E}_{w_{11}, w_{12}}(v_{2})$.\\
\indent
To simplify the analysis, we ignored two simple transformation functions $\phi(x)$ and $\psi(x)$, and also ingored the operation $\oplus 1$, without affecting the correctness of the conclusion, because $\phi(x)$ and $\psi(x)$ are two simple linear functions.  \\
\indent
From the above, the problem of cracking the key $(w_{1}$, $w_{2}$, $w_{3}$, $w_{4}$, $w_{5}$, $w_{6}$, $w_{7}$, $w_{8}$, $w_{9}$, $w_{10}$, $w_{11}$, $w_{12})$ is equivalent to solving the following system of equations. \\
\indent
\begin{equation}
\tag{5.3}
\begin{cases}
(P_{1} \oplus z_{1}, \mu_{1}) \bigotimes (\mathbb{E}_{w_{1}, w_{2}}(x_{1}), \mathbb{E}_{w_{3}, w_{4}}(y_{1})) = (C_{1}, \lambda_{1}) \\
(*, \mu_{2}) \bigotimes (\mathbb{E}_{w_{5}, w_{6}}(x_{2}), \mathbb{E}_{w_{7}, w_{8}}(y_{2})) = (C_{2}, \lambda_{2}) \\
(*, \mu_{3}) \bigotimes (\mathbb{E}_{w_{5}, w_{6}}(x_{3}), \mathbb{E}_{w_{7}, w_{8}}(y_{3})) = (C_{3}, \lambda_{3}) \\
(*, \mu_{4}) \bigotimes (\mathbb{E}_{w_{5}, w_{6}}(x_{4}), \mathbb{E}_{w_{7}, w_{8}}(y_{4})) = (C_{4}, \lambda_{4}) \\
(*, \mu_{5}) \bigotimes (\mathbb{E}_{w_{5}, w_{6}}(x_{5}), \mathbb{E}_{w_{7}, w_{8}}(y_{5})) = (C_{5}, \lambda_{5}) \\
(*, \mu_{6}) \bigotimes (\mathbb{E}_{w_{5}, w_{6}}(x_{6}), \mathbb{E}_{w_{7}, w_{8}}(y_{6})) = (C_{6}, \lambda_{6}) \\
(P_{2} \oplus z_{2}, \mu_{7}) \bigotimes (\mathbb{E}_{w_{1}, w_{2}}(x_{7}), \mathbb{E}_{w_{3}, w_{4}}(y_{7})) = (C_{7}, \lambda_{7}) \\
(*, \mu_{8}) \bigotimes (\mathbb{E}_{w_{5}, w_{6}}(x_{8}), \mathbb{E}_{w_{7}, w_{8}}(y_{8})) = (C_{8}, \lambda_{8}) \\
...... \\
(P_{K} \oplus z_{K}, \mu_{6K-5}) \bigotimes (\mathbb{E}_{w_{1}, w_{2}}(x_{6K-5}), \mathbb{E}_{w_{3}, w_{4}}(y_{6K-5})) \\
\,\,\,\,\,\,\,\,\,\,\,\,\,\,\,\,\,\,\,\,\,\,\,\,\,\,= (C_{6K-5}, \lambda_{6K-5}) \\
(*, \mu_{6K-4}) \bigotimes (\mathbb{E}_{w_{5}, w_{6}}(x_{6K-4}), \mathbb{E}_{w_{7}, w_{8}}(y_{6K-4})) \\
\,\,\,\,\,\,\,\,\,\,\,\,\,\,\,\,\,\,\,\,\,\,\,\,\,\,= (C_{6K-4}, \lambda_{6K-4}) \\
...... \\
(*, \mu_{6K-1}) \bigotimes (\mathbb{E}_{w_{5}, w_{6}}(x_{6K-1}), \mathbb{E}_{w_{7}, w_{8}}(y_{6K-1})) \\
\,\,\,\,\,\,\,\,\,\,\,\,\,\,\,\,\,\,\,\,\,\,\,\,\,\,= (C_{6K-1}, \lambda_{6K-1}) \\
(*, \mu_{6K}) \bigotimes (\mathbb{E}_{w_{5}, w_{6}}(x_{6K}), \mathbb{E}_{w_{7}, w_{8}}(y_{6K})) \\
\,\,\,\,\,\,\,\,\,\,\,\,\,\,\,\,\,\,\,\,\,\,\,\,\,\,= (C_{6K}, \lambda_{6K}) \\
(t_{6K+1}, \mu_{6K+1}) \bigotimes (\mathbb{E}_{w_{1}, w_{2}}(x_{6K+1}), \mathbb{E}_{w_{3}, w_{4}}(y_{6K+1})) \\
\,\,\,\,\,\,\,\,\,\,\,\,\,\,\,\,\,\,\,\,\,\,\,\,\,\,= (C_{6K+1}, \lambda_{6K+1}) \\
\end{cases}
\end{equation}
\begin{equation}
\tag{5.4}
\begin{cases}
\mathbb{E}_{w_{u_{1}}, w_{v_{1}}}(\lambda_{1})=\mu_{2} \\
\mathbb{E}_{w_{u_{2}}, w_{v_{2}}}(\lambda_{2})=\mu_{3} \\
\mathbb{E}_{w_{u_{2}}, w_{v_{2}}}(\lambda_{3})=\mu_{4} \\
\mathbb{E}_{w_{u_{2}}, w_{v_{2}}}(\lambda_{4})=\mu_{5} \\
\mathbb{E}_{w_{u_{2}}, w_{v_{2}}}(\lambda_{5})=\mu_{6} \\
\mathbb{E}_{w_{u_{2}}, w_{v_{2}}}(\lambda_{6})=\mu_{7} \\
...... \\
\mathbb{E}_{w_{u_{1}}, w_{v_{1}}}(\lambda_{6K-5})=\mu_{6K-4} \\
\mathbb{E}_{w_{u_{2}}, w_{v_{2}}}(\lambda_{6K-4})=\mu_{6K-3} \\
\mathbb{E}_{w_{u_{2}}, w_{v_{2}}}(\lambda_{6K-3})=\mu_{6K-2} \\
\mathbb{E}_{w_{u_{2}}, w_{v_{2}}}(\lambda_{6K-2})=\mu_{6K-1} \\
\mathbb{E}_{w_{u_{2}}, w_{v_{2}}}(\lambda_{6K-1})=\mu_{6K} \\
\mathbb{E}_{w_{u_{2}}, w_{v_{2}}}(\lambda_{6K})=\mu_{6K+1} \\
\end{cases}
\end{equation}
\begin{equation}
\tag{5.5}
\begin{cases}
\mathbb{E}_{w_{9}, w_{10}}(u_{1})=w_{u_{1}} \\
\mathbb{E}_{w_{9}, w_{10}}(v_{1})=w_{v_{1}} \\
\mathbb{E}_{w_{11}, w_{12}}(u_{2})=w_{u_{2}} \\
\mathbb{E}_{w_{11}, w_{12}}(v_{2})=w_{v_{2}} \\
\end{cases}
\end{equation}
\\
\indent
Where $(w_{1}, w_{2}, w_{3}, w_{4}, w_{5}, w_{6}, w_{7}, w_{8}, w_{9}, w_{10}, w_{11}, w_{12})$ are unknown variables to be solved. $(P_{1}$, $P_{2}$, ..., $P_{K})$, $(C_{1}$, $C_{2}$, ..., $C_{6K+1})$, $(x_{1}$, $x_{2}$,..., $x_{6K+1})$, $(y_{1}$, $y_{2}$,..., $y_{6K+1})$, $(z_{1}$, $z_{2}$,..., $z_{K})$, $u_{1}, v_{1}, u_{2}, v_{2}, t_{6K+1}$  are known variables. $(\lambda_{1}$, $\lambda_{2}$, ..., $\lambda_{6K+1})$, $(\mu_{1}$, $\mu_{2}$, ..., $\mu_{6K+1})$ are implicit variables determined by random numbers.  \\
\indent
The proof process is divided into the following five steps. \\
\indent
(i) When $K>12$, the system of equations has an unique solution or no solutions. \\
\indent
(ii) Instead of directly solving for the 12 unknowns variables $(w_{1}$, $w_{2}$, $w_{3}$, $w_{4}$, $w_{5}$, $w_{6}$, $w_{7}$, $w_{8}$, $w_{9}$, $w_{10}$, $w_{11}$, $w_{12})$, we first consider a weaker problem, verifying the correctness of the five variables $(w_{1}$, $w_{2}$, $w_{3}$, $w_{4}$, $w_{9})$, this is equivalent to solving for the other seven variables $(w_{5}$, $w_{6}$, $w_{7}$, $w_{8}$, $w_{10}$, $w_{11}$, $w_{12})$ given specific values for $(w_{1}$, $w_{2}$, $w_{3}$, $w_{4}$, $w_{9})$. In the following, we will prove that the computational complexity of verifying the five variables $(w_{1}$, $w_{2}$, $w_{3}$, $w_{4}$, $w_{9})$ is at least $O(2^{L-1})$. \\
\indent
(iii) When $(w_{1}, w_{2}, w_{3}, w_{4}, w_{9})$ given, to solve $(w_{5}$, $w_{6}$, $w_{7}$, $w_{8}$, $w_{10}$, $w_{11}$, $w_{12})$, after performing elimination on the implicit variables $\lambda_{i}, \mu_{i}$, the system of equations (5.3) (5.4) (5.5) can be equivalently transformed into a system of general quintic equations about the unknown variables $w_{11} \oplus w_{12}$ with coefficient given by $(w_{5}$, $w_{6}$, $w_{7}$, $w_{8}$, $w_{10}$, $w_{11})$. According to the common conclusion that there are no algebraic solutions for quintic equations, $w_{11} \oplus w_{12}$ can't be represented by algebraic expressions with the variables $(w_{5}$, $w_{6}$, $w_{7}$, $w_{8}$, $w_{10}$, $w_{11})$, we cannot establish a system of algebraic  equations only containing $(w_{5}$, $w_{6}$, $w_{7}$, $w_{8},$ $w_{10}$, $w_{11})$ without $w_{11} \oplus w_{12}$, thus we can't solve the variables $(w_{5}, w_{6}, w_{7}, w_{8}, w_{10}, w_{11})$ without  $w_{11} \oplus w_{12}$. \\
\indent
(iv) When $(w_{1}$, $w_{2}$, $w_{3}$, $w_{4}$, $w_{9})$ given, without solving for $(w_{5}$, $w_{6}$, $w_{7}$, $w_{8}$, $w_{11}$, $w_{12})$, we will prove that the problem of solving $(w_{10})$ is equivalent to the problem of solving the slope determined by $w_{9} \oplus w_{10}$ of a line given only the x-coordinates of several points on it. Obviously there is no universal algorithm to solve it without brute-force searching any $\mu_{i}$ or $w_{11} \oplus w_{12}$. \\
\indent
(v) Together with (ii) (iii) and (iv), we will prove that the computational complexity of the problem of solving (5.3), (5.4), (5.5) is at least $O(2^{L-1})$. \\
\indent
Now we prove (i). \\
\indent
When $K>12$, the number of constraints in the system of equations (5.3), (5.4), (5.5) exceeds the number of unknown variables, which means that the system of equations typically has only one solution or no solutions. This completes the proof of (i).  \\
\indent
Now we prove (ii). \\
\indent
In this proof, we do not intend to directly discuss the computational complexity of solving $(w_{1}$, $w_{2}$, $w_{3}$, $w_{4}$, $w_{5}$, $w_{6}$, $w_{7}$, $w_{8}$, $w_{9}$, $w_{10}$, $w_{11}$, $w_{12})$. We only need to discuss the computational complexity of a weaker problem, which is to verify the correctness of the five variables $(w_{1}$, $w_{2}$, $w_{3}$, $w_{4}$, $w_{9})$. As long as the computational complexity of verifying the correctness of the five variables $(w_{1}, w_{2}, w_{3}, w_{4}, w_{9})$ is at least $O(2^{L-1})$, the computational complexity of solving $(w_{1}$, $w_{2}$, $w_{3}$, $w_{4}$, $w_{5}$, $w_{6}$, $w_{7}$, $w_{8}$, $w_{9}$, $w_{10}$, $w_{11}$, $w_{12})$ must not be lower than $O(2^{L-1})$. This will be explained in detail in (v). \\
\indent
From the equations (5.3) (5.4) (5.5), it is not difficult to see that the problem of verifying the correctness of $(w_{1}, w_{2}, w_{3}, w_{4}, w_{9})$ is equivalent to solve the remaining seven variables $(w_{5}, w_{6}, w_{7}, w_{8}, w_{9}, w_{10}, w_{11}, w_{12})$ when given a set of specific values for the four variables $(w_{1}, w_{2}, w_{3}, w_{4}, w_{9})$.\\
\indent
Now we prove (iii). \\
\indent
Perform a simple variable substitution for $w_{5}, w_{6}, w_{7}, w_{8}, w_{9}, w_{10}, w_{11}, w_{12}$ as $\widetilde{\delta_{5}}=w_{5} \oplus w_{6}$, $\widetilde{\delta_{7}}=w_{7} \oplus w_{8}$, $\widetilde{\delta_{9}}=w_{9} \oplus w_{10}$, $\widetilde{\delta_{11}}=w_{11} \oplus w_{12}$, $\widetilde{w_{5}}=\theta(w_{5})=w_{5}^{+{(L-1)}}$, $\widetilde{w_{7}}=\theta(w_{7})=w_{7}^{+{(L-1)}}$, $\widetilde{w_{9}}=\theta(w_{9})=w_{9}^{+{(L-1)}}$, $\widetilde{w_{11}}=\theta(w_{11})=w_{11}^{+{(L-1)}}$. Additionally we denote $\widetilde{P_{i}}=P_{i} \oplus z_{i}$. Then for any  $i \not \equiv 1 \pmod{6}$ we have \\
\begin{equation}
\tag{5.6}
\begin{cases}
w_{x_{i}}=\mathbb{E}_{w_{5}, w_{6}}(x_{i})=x_{i} * \widetilde{\delta_{5}} \oplus \widetilde{w_{5}} \\
w_{y_{i}}=\mathbb{E}_{w_{7}, w_{8}}(y_{i})=y_{i} * \widetilde{\delta_{7}} \oplus \widetilde{w_{7}} \\
w_{u_{1}}=\mathbb{E}_{w_{9}, w_{10}}(u_{1})=u_{1} * \widetilde{\delta_{9}} \oplus \widetilde{w_{9}} \\
w_{v_{1}}=\mathbb{E}_{w_{9}, w_{10}}(v_{1})=v_{1} * \widetilde{\delta_{9}} \oplus \widetilde{w_{9}} \\
w_{u_{2}}=\mathbb{E}_{w_{11}, w_{12}}(u_{2})=u_{2} * \widetilde{\delta_{11}} \oplus \widetilde{w_{11}} \\
w_{v_{2}}=\mathbb{E}_{w_{11}, w_{12}}(v_{2})=v_{2} * \widetilde{\delta_{11}} \oplus \widetilde{w_{11}}
\end{cases}
\end{equation}
\indent
When give $(w_{1}, w_{2}, w_{3}, w_{4})$, $w_{x_{i}}$($i \equiv 1 \pmod{6}$) can be calculated by $w_{x_{i}}=\mathbb{E}_{w_{1}, w_{2}}(x_{i})$ and $w_{y_{i}}$($y \equiv 1 \pmod{6}$) can be calculated by $w_{y_{i}}=\mathbb{E}_{w_{3}, w_{4}}(y_{i})$, then $\lambda_{1}$ can be directly calculated from the first equation of (5.3) as \\
\begin{equation}
\lambda_{1}= \widetilde{P_{1}} * (w_{x_{1}} \oplus w_{y_{1}}) \oplus \theta(w_{x_{1}})  \notag
\end{equation}
\indent
Since $\mu_{2}=\mathbb{E}_{w_{u_{1}}, w_{v_{1}}}(\lambda_{1})$ from the first equation of (5.4), $\mu_{2}$ can be expressed as \\
\begin{equation}
\notag
\begin{aligned}
\mu_{2}=\lambda_{1} * (w_{u_{1}} \oplus w_{v_{1}}) \oplus \theta(w_{u_{1}}) \\
=\lambda_{1} * (u_{1} * \widetilde{\delta_{9}} \oplus v_{1} * \widetilde{\delta_{9}}) \oplus \theta(u_{1} * \widetilde{\delta_{9}} \oplus \widetilde{w_{9}}) 
\end{aligned}
\end{equation}
\indent
Since $\lambda_{2}=\mu_{2}*(w_{x_{2}} \oplus w_{y_{2}}) \oplus C_{2}$ from the second equation of (5.3), $\lambda_{2}$ can be expressed as \\
\begin{equation}
\notag
\begin{aligned}
\lambda_{2}=(\lambda_{1} * (u_{1} * \widetilde{\delta_{9}} \oplus v_{1} * \widetilde{\delta_{9}}) \oplus \theta(u_{1} * \widetilde{\delta_{9}} \oplus \widetilde{w_{9}})) * (\mathbb{E}_{w_{5}, w_{6}}(x_{2}) \oplus \mathbb{E}_{w_{7}, w_{8}}(y_{2})) \\
\oplus C_{2}
\end{aligned}
\end{equation}
\indent
Similar to the step-by-step process above, we can obtain the expressions for $\mu_{3}$, $\lambda_{3}$, $\mu_{4}$, $\lambda_{4}$, $\mu_{5}$, $\lambda_{5}$, $\mu_{6}$, $\lambda_{6}$, $\mu_{7}$, $\lambda_{7}$ separately, which $\lambda_{7}$ is as follow \\
\begin{equation}
\tag{5.7}
\begin{aligned}
\lambda_{7}= \\
(((((((((((\lambda_{1} * (u_{1} * \widetilde{\delta_{9}} \oplus v_{1} * \widetilde{\delta_{9}}) \oplus \theta(u_{1} * \widetilde{\delta_{9}} \oplus \widetilde{w_{9}})) * (\mathbb{E}_{w_{5}, w_{6}}(x_{2}) \oplus \mathbb{E}_{w_{7}, w_{8}}(y_{2})) \\
\oplus C_{2}) * (u_{2} * \widetilde{\delta_{11}} \oplus v_{2} * \widetilde{\delta_{11}}) \oplus \theta(u_{2} * \widetilde{\delta_{11}} \oplus \widetilde{w_{11}})) * (\mathbb{E}_{w_{5}, w_{6}}(x_{3}) \oplus \mathbb{E}_{w_{7}, w_{8}}(y_{3}))  \\
\oplus C_{3}) * (u_{2} * \widetilde{\delta_{11}} \oplus v_{2} * \widetilde{\delta_{11}}) \oplus \theta(u_{2} * \widetilde{\delta_{11}} \oplus \widetilde{w_{11}})) * (\mathbb{E}_{w_{5}, w_{6}}(x_{4}) \oplus \mathbb{E}_{w_{7}, w_{8}}(y_{4}))  \\
\oplus C_{4}) * (u_{2} * \widetilde{\delta_{11}} \oplus v_{2} * \widetilde{\delta_{11}}) \oplus \theta(u_{2} * \widetilde{\delta_{11}} \oplus \widetilde{w_{11}})) * (\mathbb{E}_{w_{5}, w_{6}}(x_{5}) \oplus \mathbb{E}_{w_{7}, w_{8}}(y_{5}))  \\
\oplus C_{5}) * (u_{2} * \widetilde{\delta_{11}} \oplus v_{2} * \widetilde{\delta_{11}}) \oplus \theta(u_{2} * \widetilde{\delta_{11}} \oplus \widetilde{w_{11}})) * (\mathbb{E}_{w_{5}, w_{6}}(x_{6}) \oplus \mathbb{E}_{w_{7}, w_{8}}(y_{6}))  \\
\oplus C_{6}) * (u_{2} * \widetilde{\delta_{11}} \oplus v_{2} * \widetilde{\delta_{11}}) \oplus \theta(u_{2} * \widetilde{\delta_{11}} \oplus \widetilde{w_{11}})) * (w_{x_{7}} \oplus w_{y_{7}})   \\
\oplus C_{7}
\end{aligned}
\end{equation}
\indent
On the other hand, $\lambda_{7}$ can be directly calculated as  \\
\begin{equation}
\lambda_{7}=\widetilde{P_{2}} * (w_{x_{7}} \oplus w_{y_{7}}) \oplus \theta(w_{x_{7}})  \tag{5.8}
\end{equation}
\indent
Together with (5.7) and (5.8), we obtain the following equation \\
\begin{equation}
\tag{5.9}
\begin{aligned}
(((((((((((\lambda_{1} * (u_{1} * \widetilde{\delta_{9}} \oplus v_{1} * \widetilde{\delta_{9}}) \oplus \theta(u_{1} * \widetilde{\delta_{9}} \oplus \widetilde{w_{9}})) * (\mathbb{E}_{w_{5}, w_{6}}(x_{2}) \oplus \mathbb{E}_{w_{7}, w_{8}}(y_{2})) \\
\oplus C_{2}) * (u_{2} * \widetilde{\delta_{11}} \oplus v_{2} * \widetilde{\delta_{11}}) \oplus \theta(u_{2} * \widetilde{\delta_{11}} \oplus \widetilde{w_{11}})) * (\mathbb{E}_{w_{5}, w_{6}}(x_{3}) \oplus \mathbb{E}_{w_{7}, w_{8}}(y_{3}))  \\
\oplus C_{3}) * (u_{2} * \widetilde{\delta_{11}} \oplus v_{2} * \widetilde{\delta_{11}}) \oplus \theta(u_{2} * \widetilde{\delta_{11}} \oplus \widetilde{w_{11}})) * (\mathbb{E}_{w_{5}, w_{6}}(x_{4}) \oplus \mathbb{E}_{w_{7}, w_{8}}(y_{4}))  \\
\oplus C_{4}) * (u_{2} * \widetilde{\delta_{11}} \oplus v_{2} * \widetilde{\delta_{11}}) \oplus \theta(u_{2} * \widetilde{\delta_{11}} \oplus \widetilde{w_{11}})) * (\mathbb{E}_{w_{5}, w_{6}}(x_{5}) \oplus \mathbb{E}_{w_{7}, w_{8}}(y_{5}))  \\
\oplus C_{5}) * (u_{2} * \widetilde{\delta_{11}} \oplus v_{2} * \widetilde{\delta_{11}}) \oplus \theta(u_{2} * \widetilde{\delta_{11}} \oplus \widetilde{w_{11}})) * (\mathbb{E}_{w_{5}, w_{6}}(x_{6}) \oplus \mathbb{E}_{w_{7}, w_{8}}(y_{6}))  \\
\oplus C_{6}) * (u_{2} * \widetilde{\delta_{11}} \oplus v_{2} * \widetilde{\delta_{11}}) \oplus \theta(u_{2} * \widetilde{\delta_{11}} \oplus \widetilde{w_{11}})) * (w_{x_{7}} \oplus w_{y_{7}})   \\
\oplus C_{7} \\
=\widetilde{P_{2}} * (w_{x_{7}} \oplus w_{y_{7}}) \oplus \theta(w_{x_{7}})
\end{aligned}
\end{equation}
\indent
Where $\widetilde{\delta_{9}}, \widetilde{\delta_{11}}, \widetilde{w_{11}}, w_{5}, w_{6}, w_{7}, w_{8}$ are unknown variables, others are known variables. \\
\indent
It is not difficult to find that (5.9) is a general quintic equation about the variable $\widetilde{\delta_{11}}$ with coefficient as given by $(\widetilde{\delta{9}}, \widetilde{w_{11}}, w_{5}, w_{6}, w_{7}, w_{8})$. This is because when (5.9) is expanded into an expression connected by $\oplus$, the continuous right multiplication operations on $\widetilde{\delta_{11}}$ are executed at most five times. \\
\indent
In fact, the above explanation is reasonable. This is because the $\oplus$ can be viewed as traditional $+$, the right multiplication can be viewed as traditional $\times$, the repeated right multiplication of the same parameter can be viewed as a power operation with respect to that parameter. Similarly, radical operations based on right multiplication can also be well defined. Thus, the conclusion that the general quintic equation has no algebraic solutions can be entirely extended to the general quintic equation concerning right multiplication. For more details, please refer to Appendix A. \\
\indent
Similar to the elimination process in (5.9), after performing elimination on $\lambda_{6i+1}$, $\mu_{6i+2}$, $\lambda_{6i+2}$, $\mu_{6i+3}$, $\lambda_{6i+3}$, $\mu_{6i+4}$, $\lambda_{6i+4}$, $\mu_{6i+5}$, $\lambda_{6i+5}$, $\mu_{6i+6}$, $\lambda_{6i+6}$, $\mu_{6i+7}$, $\lambda_{6i+7}$, more general quintic equaions about the variables $\widetilde{\delta_{11}}$ can be established with coefficient as given by $(\widetilde{\delta{9}}, \widetilde{w_{11}}, w_{5}, w_{6}, w_{7}, w_{8})$. \\
\indent
That is to say, after performing elimination operations on all implicit variables $\lambda_{i}, \mu_{i}$, the system of equations (5.3) (5.4) (5.5) are equivalently transformed into some quintic equations about the variable $\widetilde{\delta_{11}}$ with coefficient as given by $(\widetilde{\delta{9}}, \widetilde{w_{11}}, w_{5}, w_{6}, w_{7}, w_{8})$. \\
\indent
According to the common conclusion that there are no algebraic solutions for quintic equations, $\widetilde{\delta_{11}}=w_{11} \oplus w_{12}$ can't be represented by any algebraic expression with the six variables $(\widetilde{\delta{9}}, \widetilde{w_{11}}, w_{5}, w_{6}, w_{7}, w_{8})$, thus we can't establish a system of equations only containing $(\widetilde{\delta{9}}, \widetilde{w_{11}}, w_{5}, w_{6}, w_{7}, w_{8})$ without $\widetilde{\delta_{11}}$. This completes the proof of (iii). \\
\indent
It is noteworthy that the aforementioned quintic equations not only lacks an algebraic solution but also cannot be solved by any approximate numerical method. This is because, based on the right multiplication operation constructed in this paper, it is impossible to establish an ``order" relation $\prec$, such that for any $\delta_{1} \prec \delta_{2}$,  $m * \delta_{1} \prec m * \delta_{2}$ holds for all $m$, please refer to Appendix A for details of the proof. Thus it becomes impossible to establish any numerical simulation on the basis of this right multiplication operation. \\
\indent
Now we prove (iv).  \\
\indent
When given $(w_{1}, w_{2}, w_{3}, w_{4}, w_{9})$, to solve $\widetilde{\delta_{9}}=w_{9} \oplus w_{10}$, we need to analyze the implicit variables $\lambda_{i}, \mu_{i}$. From (5.5), since $w_{u_{1}}=\mathbb{E}_{w_{9}, w_{10}}(u_{1})$ and $w_{v_{2}}=\mathbb{E}_{w_{9}, w_{10}}(v_{2})$, the problem of solving $\widetilde{\delta_{9}}$ is equivalent to the problem of solving $(w_{u_{1}}, w_{v_{1}})$. \\
\indent
Perform a simple variable substitution for $w_{u_{1}}, w_{v_{1}}$ as $\widetilde{u}=w_{u_{1}} \oplus w_{v_{1}}$ and $\widetilde{v}=w_{u_{1}}^{+(L-1)}$, $(\lambda_{i}, \mu_{i+1})$($i \equiv 1 \mod 6$) is a set of points that satisfy the linear equation $\mu=\lambda * \widetilde{u} \oplus \widetilde{v}$, where $\widetilde{u},\widetilde{v}$ are parameters, the system of equations (5.4) can be represented by the following figure. \\
\begin{tikzcd}
                 & \mu              &                                                             &                               &                                 &         \\
                 &                  &                                                             &                               & {p_{3}=(\lambda_{13},\mu_{14})} &         \\
                 &                  &                                                             & {p_{2}=(\lambda_{7},\mu_{8})} &                                 &         \\
                 &                  & {p_{1}=(\lambda_{1},\mu_{2})} \arrow[rruu, no head, dotted] &                               &                                 &         \\
{} \arrow[rrrrr] &                  &                                                             &                               &                                 & \lambda \\
                 & {} \arrow[uuuuu] &                                                             &                               &                                 &        
\end{tikzcd}
\\
\indent
In fact, the straight line in the above figure does not represent a true straight line, it is just an expression of a linear relationship between $\mu$ and $\lambda$.  \\
\indent
When given $(w_{1}, w_{2}, w_{3}, w_{4})$, we can calculate $\lambda_{6i+1}$ and $\mu_{6i+1}$ $(1 \le i \le K)$ from the $6i+1$-th equation in (5.3) directly. \\
\indent
According to the (iii), because $(\widetilde{\delta{9}}, \widetilde{w_{11}}, w_{5}, w_{6}, w_{7}, w_{8})$ can't be solved , must be expressed with at least one free variable $\widetilde{\delta_{11}}$, thus $\mu_{2}$, $\mu_{8}$, ..., $\mu_{6K+2}$ can't be solved without brute-force searching for $\widetilde{\delta_{11}}$. \\
\indent  
The problem of solving $\widetilde{\delta_{9}}$ when given $(w_{1}, w_{2}, w_{3}, w_{4}, w_{9})$ is equivalent to the problem of solving $(\widetilde{u})$ when only $\lambda_{6i+1}$ $(1 \le i \le K)$ known, where $\lambda_{6i+1}$ is the horizontal coordinate of the $i$-th points of the linear equation $\mu=\lambda * \widetilde{u} \oplus \widetilde{v}$. \\
\indent
Obviously, there is no universal algorithm to solve $\widetilde{u}=(u_{1} \oplus v_{1}) * ( w_{9} \oplus w_{10})$ without brute-force searching any $\mu_{6i+2}$ or $\widetilde{\delta_{11}}$. \\
\indent
This completes the proof of (iv). \\
\indent
Now we prove (v). \\
\indent
Returning to the problem of verifying $(w_{1}, w_{2}, w_{3}, w_{4}, w_{9})$, we have proved that there is no universal algorithm except for brute-force searching at least one variable. \\
\indent
Since the computational complexity of brute-force searching for $\widetilde{\delta_{11}}=w_{11} \oplus w_{12}$ or $\mu_{6i+2}$ is $O(2^{L-1})$, thus the complexity of verifying $(w_{1}, w_{2}, w_{3}, w_{4}, w_{9})$ is at least $O(2^{L-1})$. \\
\indent
Obviously, if we solved $(w_{1}$, $w_{2}$, $w_{3}$, $w_{4}$, $w_{5}$, $w_{6}$, $w_{7}$, $w_{8}$, $w_{9}$, $w_{10}$, $w_{11}$, $w_{12})$, we can immediately solve $(w_{1}$, $w_{2}$, $w_{3}$, $w_{4}$, $w_{9}$), thus  the computational complexity of solving $(w_{1}$, $w_{2}$, $w_{3}$, $w_{4}$, $w_{5}$, $w_{6}$, $w_{7}$, $w_{8}$, $w_{9}$, $w_{10}$, $w_{11}$, $w_{12})$ is not lower than the computational complexity of solving $(w_{1}$, $w_{2}$, $w_{3}$, $w_{4}$, $w_{9}$). \\
\indent
Since the complexity of verifying a solution should not be higher than solving it, the computational complexity of solving $(w_{1}$, $w_{2}$, $w_{3}$, $w_{4}$, $w_{9}$) is not lower than the computational complexity of verifying $(w_{1}$, $w_{2}$, $w_{3}$, $w_{4}$, $w_{9}$). So, the computational complexity of solving the system of equations satisfies the following relationship \\
\begin{equation}
\notag
\begin{aligned}
Complexity(Solve (w_{1},w_{2},w_{3},w_{4},w_{5},w_{6},w_{7},w_{8},w_{9},w_{10},w_{11},w_{12})) \\
 \geq Complexity(Solve(w_{1},w_{2}, w_{3}, w_{4},w_{9})) \\
 \geq Complexity(Verify(w_{1},w_{2}, w_{3}, w_{4},w_{9})) \\
 \geq O(2^{L-1}) \\
\end{aligned} 
\end{equation}
\indent
In summary, in the process of solving the system of equations, the computational complexity of solving $(w_{1}$, $w_{2}$, $w_{3}$, $w_{4}$, $w_{5}$, $w_{6}$, $w_{7}$, $w_{8}$, $w_{9}$, $w_{10}$, $w_{11}$, $w_{12})$ is at least $O(2^{L-1})$. \\
\indent
\end{proof}

\indent
According to the conclusion of Theorem 12, the computational complexity of solving $f^{-1}$ is exponential. \\
\indent
Due to the fact that solving $f$ can be completed in polynomial time and the computational complexity of solving $f^{-1}$ is exponential, $f$ is a one-way function, we can summarize it to the following theorem. \\
\\
\indent
\textbf{[Theorem 13]}  The function $f$ defined in (5.1) is a one-way function.\\
\\
\indent
According to Theorem 11 and Theorem 13, we conclude that P$\neq$NP.

\section{How to achieve only exhaustive search to crack keys}
\indent
Without causing ambiguity, we denote the operation of performing two right multiplications by $\delta$ on $R$ as $R * \delta^{2}$.\\
\indent
Similarly, we can use $S / \delta^{2}$ to represent performing two division operations on $S$ with respect to $\delta$. \\
\indent
Now we give the two algorithms $m * \delta^{2}$ and $s /\delta^{2}$ as follows\\
\begin{algorithm} [H]
\caption{calculation of $m * \delta^{2}$.}
\label{alg: Framework}
\begin{algorithmic}[1]
\REQUIRE ~~ \\
Input: $m, \delta$. \\
\ENSURE ~~ \\
Output: $m * \delta^{2}$. \\
\STATE $m \leftarrow \phi(m)$. \\
\STATE $s \leftarrow m * \delta$. \\
\STATE $s \leftarrow \phi(s)$. \\
\STATE $s \leftarrow s * \delta$. \\
\RETURN $s$. \\
\end{algorithmic}
\end{algorithm}
\indent
Where $m$ and $\delta$ both contain odd number of bits that are 1, the output $s$ also contains odd number of bits that are 1. \\

\begin{algorithm} [H]
\caption{calculation of $s / \delta^{2}$.}
\label{alg: Framework}
\begin{algorithmic}[1]
\REQUIRE ~~ \\
Input: $s, \delta$. \\
\ENSURE ~~ \\
Output: $s / \delta^{2}$. \\
\STATE $m \leftarrow s / \delta$. \\
\STATE $m \leftarrow \psi(m)$. \\
\STATE $m \leftarrow m / \delta$. \\
\STATE $m \leftarrow \psi(m).$ \\
\RETURN $m$. \\
\end{algorithmic}
\end{algorithm}
\indent
Where $s$ and $\delta$ both contain odd number of bits that are 1, the output $m$ also contains odd number of bits that are 1. \\

\indent
\subsection{Design of encryption algorithm for only exhaustive method to crack the key}
\indent
Although we found an encryption algorithm in Chapter 5 that, given the known correspondence between plaintext and ciphertext, the computational complexity for cracking the key is at least exponentially to the length of key as $O(2^{L-1})$, this does not mean that this algorithm is practical. Because the length of the key $(w_{1}$, $w_{2}$, $w_{3}$, $w_{4}$, $w_{5}$, $w_{6}$, $w_{7}$, $w_{8}$, $w_{9}$, $w_{10}$, $w_{11}$, $w_{12})$ is $12L-12$ bits, theoretically, the computational  complexity for cracking the key must reach $O(2^{12L-12})$ to recognize its secure.\\
\indent
In this chapter, we utilized the Eagle encoding algorithm and its related features to design another short key symmetric block encryption algorithm, which we named the $Eagle^{*}$ encryption algorithm. The main feature of this algorithm is that it introduce a nonlinear operation, together with the random numbers as designed in Eagle encryption algorithm. \\
\indent
After being encrypted by this encryption algorithm, the problem of cracking the keys given that all plaintext ciphertext correspondences are known, is equivalent to solving a system of nonlinear equations. \\
\indent
In addition, the key is consists of three variables, for verifying any two variables in them needs to solving a system of quintic equations,  according to the quintic equation with no algebraic solution, we assert that there is no fast algorithm to verify any two variables, and thus verifying any two variables requires exhaustive enumeration of the other variable. So to crack the key, it is necessary to exhaustively search for all values in these three variables.\\
\indent
The entire $Eagle^{*}$ encryption algorithm is divided into three parts: key generation, encryption algorithm, and decryption algorithm. Next, we will provide a detailed description.\\

\indent
\subsubsection{$Eagle^{*}$ key generator}
\indent
The key is divided into three independent variables, denoted as $s_{x}$, $s_{y}$, $s_{t}$, where $s_{x}$, $s_{y}$, $s_{t}$ all contain odd number of bits that are 1. \\
\indent
Assuming $s_{x}, s_{y}, s_{t}$ are all $L$ bits, the length of the entire key can be considered as $3L-3$ bits, this is because each variable has a loss of 1 bit. \\

\subsubsection{$Eagle^{*}$ encryption algorithm}
\indent
The plaintext $P$ is grouped by $L-1$ bits, and the last group with less than $L-1$ bits are randomly filled into $L-1$ bits, then fill one bit with 1 to the last bit of each group to ensure the length of each group is $L$ bits. The total number of groups is assumed to be $K$, denoted as $P=(P_{1}, P_{2}, ..., P_{K})$. Obviously, $P_{i}$ are all odd numbers. \\
\indent
We randomly generate $6K+3$ groups number with each have $L$ bits, denoted as $\delta x=(\delta x_{1}$, $\delta x_{2}$, ..., $\delta x_{2K+1})$, $\delta y=(\delta y_{1}$, $\delta y_{2}$, ..., $\delta y_{2K+1})$, $\delta t=(\delta t_{1}$, $\delta t_{2}$, ..., $\delta t_{2K+1})$, where $\delta x_{i}$ contains odd number of bits that are 1, $\delta y_{i}$ contains odd number of bits that are 1, $\delta t_{i}$ contains odd number of bits that are 1. \\
\indent
Then we randomly generate three number as $\delta u, \delta v$, $M_{0}$, where $\delta u, \delta v, M_{0}$ all contain odd number of bits that are 1.\\
\indent
The encryption process is detailed in Algorithm 11 below. \\
\begin{algorithm} [H]
\caption{$Eagle^{*}$-Encrypt.}
\label{alg: Framework}
\begin{algorithmic}[1]
\REQUIRE ~~ \\
Keys: $(s_{x}, s_{y}, s_{t})$. \\
Plaintext sequences: $P=(P_{1}, P_{2}, ..., P_{K})$. \\
Random number sequences: \\
\,\,\,\,\,\,\,\,\,\,\,\,\,\,\,\,$\delta x=(\delta x_{1}, \delta x_{2}, ..., \delta x_{2K+1})$,\\
\,\,\,\,\,\,\,\,\,\,\,\,\,\,\,\,$\delta y=(\delta y_{1}, \delta y_{2}, ..., \delta y_{2K+1})$.\\
\,\,\,\,\,\,\,\,\,\,\,\,\,\,\,\,$\delta t=(\delta t_{1}, \delta t_{2}, ..., \delta t_{2K+1})$.\\
Random number: $\delta u, \delta v, M_{0}$. \\
\ENSURE ~~ \\
Ciphertext sequences: \\
\,\,\,\,\,\,\,\,\,\,\,\,\,\,\,\,$(\delta x_{1}, \delta x_{2}, ..., \delta x_{2K+1})$, \\
\,\,\,\,\,\,\,\,\,\,\,\,\,\,\,\,$(\delta y_{1}, \delta y_{2}, ..., \delta y_{2K+1})$, \\
\,\,\,\,\,\,\,\,\,\,\,\,\,\,\,\,$(C_{1}, C_{2}, ..., C_{2K+1})$, \\
\,\,\,\,\,\,\,\,\,\,\,\,\,\,\,\,$(z_{1}, z_{2}, ..., z_{K})$, \\
\,\,\,\,\,\,\,\,\,\,\,\,\,\,\,\,$\delta u, \delta v, \delta t_{2K+1}$ \\
\STATE $u \leftarrow (s_{x} \oplus s_{y} \oplus 1) / \delta_{u}^{2}$. \\
\STATE $v \leftarrow s_{t} / \delta_{v}^{2}$. \\
\FOR{$i=1; i \le 2K+1; i++$}
	\STATE $x_{i} \leftarrow s_{x} / \delta x_{i}^{2}$. \\
	\STATE $y_{i} \leftarrow s_{y} / \delta y_{i}^{2}$. \\
\ENDFOR
\FOR{$j=1; j \le K; j++$}
	\STATE $z_{j} \leftarrow P_{j} \oplus (s_{t} / \delta t_{2j-1}^{2})$. \\
\ENDFOR
\STATE $M \leftarrow M_{0}$ \\
\STATE $C \leftarrow 0$ \\
\FOR{$k=1; k \le 2K+1; k++$}
	\STATE $(C_{k}, C) \leftarrow (\phi(\delta t_{k}), \phi(M)) \bigotimes (x_{k} \oplus 1, y_{k})$ \\
	\STATE $C \leftarrow \phi(C)$ \\
	\STATE $M \leftarrow \mathbb{E}_{u \oplus 1, v}(C)$ \\
\ENDFOR
\end{algorithmic}
\end{algorithm}

\indent
From the above algorithm, it can be seen that the $Eagle^{*}$ encryption algorithm is divided into two independent parts, as shown in the following figure, which ignored two functions $\phi(x), \psi(x)$ and the operation $\oplus 1$ in the figure. \\

\begin{tikzcd}
M_{0} \arrow[dd, "\delta t_{1}"]                                          &                                                                                         &    &                                                                                 \\
                                                                          & {} \arrow[r, "{x_{1}=s_{x}/\delta x_{1}^{2},y_{2}=s_{y}/\delta y_{1}^{2}}"]             & {} & {(\delta t_{1},M_{0})\bigotimes (x_{1}, y_{1})} \arrow[lld] \arrow[llld]        \\
{M_{1}\leftarrow \mathbb{E}_{u,v}(M_{1})} \arrow[dd, "\delta t_{2}"]      & C_{1}                                                                                   &    &                                                                                 \\
                                                                          & {} \arrow[r, "{x_{2}=s_{x}/\delta x_{2}^{2},y_{2}=s_{y}/\delta y_{2}^{2}}"]             & {} & {(\delta t_{2}, M_{1})\bigotimes (x_{2}, y_{2})} \arrow[llld] \arrow[lld]       \\
{M_{2}\leftarrow \mathbb{E}_{u,v}(M_{2})} \arrow[dd, "......"]            & C_{2}                                                                                   &    &                                                                                 \\
                                                                          &                                                                                         &    &                                                                                 \\
{M_{2K}\leftarrow \mathbb{E}_{u,v}(M_{2K})} \arrow[dd, "\delta t_{2K+1}"] & C_{K}                                                                                   &    &                                                                                 \\
                                                                          & {} \arrow[r, "{x_{2K+1}=s_{x}/\delta x_{2K+1}^{2},y_{2K+1}=s_{y}/\delta y_{2K+1}^{2}}"] & {} & {(\delta t_{2K+1},M_{2K})\bigotimes (x_{K+1},y_{k+1})} \arrow[llld] \arrow[lld] \\
M_{2K+1}                                                                  & C_{2K+1}                                                                                &    &                                                                                
\end{tikzcd}
\\
\\
\indent
As shown from the above figure, the ciphertext length is about three times the plaintext length.\\

\subsubsection{$Eagle^{*}$ decryption algorithm}
\indent
For $Eagle^{*}$ decryptors, the ciphertext sequences are $C=($$C_{1}$, $C_{2}$, ..., $C_{2K+1}$, $\delta x_{1}$, $\delta x_{2}$, ..., $\delta x_{2K+1}$, $\delta y_{1}$, $\delta y_{2}$, ..., $\delta y_{2K+1}$, $z_{1}$, $z_{2}$, ..., $z_{K})$, $\delta u$, $\delta v$ and $\delta t_{2K+1}$. Here, the ciphertext sequences can be divided into three independent parts $(C_{1}$, $C_{2}$, ..., $C_{2K+1}$, $z_{1}$, $z_{2}$, ..., $z_{K}$, $\delta t_{2K+1})$ and $(\delta x_{1}$, $\delta x_{2}$, ..., $\delta x_{2K+1})$ and $(\delta y_{1}$, $\delta y_{2}$, ..., $\delta y_{2K+1})$,  First we use $(s_{x})$ and $(\delta x_{1}$, $\delta x_{2}$, ..., $\delta x_{2K+1})$ to calculate the random number sequences $x=(x_{1}$, $x_{2}$, ..., $x_{2K+1})$, then we use $(s_{y})$ and $(\delta y_{1}$, $\delta y_{2}$, ..., $\delta y_{2K+1})$ to calculate the random number sequences $y=(y_{1}$, $y_{2}$, ..., $y_{2K+1})$, then we use the random number $u=(s_{x} \oplus s_{y} \oplus 1) / \delta_{u}^{2}$, $v=s_{t} / \delta_{v}^{2}$ and the ciphertexts $(C_{1}$, $C_{2}$, ..., $C_{2K+1}$, $z_{1}, z_{2}, ..., z_{K}, \delta t_{2K+1})$ and the random number sequences $x=(x_{1}$, $x_{2}$, ..., $x_{2K+1})$ and $y=(y_{1}$, $y_{2}$, ..., $y_{2K+1})$ to decrypt $P=(P_{1}$, $P_{2}$, ..., $P_{K})$. The specific process is detailed in the following algorithm. \\

\begin{algorithm} [H]
\caption{$Eagle^{*}$-Decrypt.}
\label{alg: Framework}
\begin{algorithmic}[1]
\REQUIRE ~~ \\
Key: $(s_{x}, s_{y}, s_{t})$. \\
Ciphertext sequences: \\
\,\,\,\,\,\,\,\,\,\,\,\,\,\,\,\,$(\delta x_{1}, \delta x_{2}, ..., \delta x_{2K+1})$, \\
\,\,\,\,\,\,\,\,\,\,\,\,\,\,\,\,$(\delta y_{1}, \delta y_{2}, ..., \delta y_{2K+1})$, \\
\,\,\,\,\,\,\,\,\,\,\,\,\,\,\,\,$(C_{1}, C_{2}, ..., C_{2K+1})$, \\
\,\,\,\,\,\,\,\,\,\,\,\,\,\,\,\,$(z_{1}, z_{2}, ..., z_{K})$, \\
\,\,\,\,\,\,\,\,\,\,\,\,\,\,\,\,$\delta u, \delta v, \delta t_{2K+1}$. \\
\ENSURE ~~ \\
Plaintext sequences: $P=(P_{1}, P_{2}, ..., P_{K})$. \\
\STATE $u \leftarrow (s_{x} \oplus s_{y} \oplus 1) / \delta_{u}^{2}$. \\
\STATE $v \leftarrow s_{t} / \delta_{v}^{2}$. \\
\FOR{$i=1; i \le 2K+1; i++$}
	\STATE $x_{i} \leftarrow s_{x} / \delta x_{i}^{2}$. \\
	\STATE $y_{i} \leftarrow s_{y} / \delta y_{i}^{2}$. \\
\ENDFOR
\STATE $M \leftarrow 0, C \leftarrow 0$ \\
\STATE $C \leftarrow \mathbb{E}_{x_{2K+1} \oplus 1, y_{2K+1}}(\phi(\delta t_{2K+1}))$. \\
\STATE $M \leftarrow (C_{2K+1} \oplus C) / (x_{2K+1} \oplus y_{2K+1} \oplus 1)$ \\
\FOR{$j=2K; j \ge 1; j--$}
	\STATE $M \leftarrow \psi(M)$ \\
	\STATE $C \leftarrow \mathbb{D}_{u \oplus 1,v}(M)$ \\
	\STATE $C \leftarrow \psi(C)$ \\
	\STATE $\delta t_{j} \leftarrow \mathbb{D}_{x_{j} \oplus 1, y_{j}}(C)$ \\
	\STATE $\delta t_{j} \leftarrow \psi(\delta t_{j})$. \\
	\STATE $M \leftarrow (C_{j} \oplus C) / (x_{j} \oplus y_{j} \oplus 1)$. \\
\ENDFOR
\FOR{$k=1; k \le K; k++$}
	\STATE $P_{k} \leftarrow z_{k} \oplus (s_{t} / \delta t_{2k-1}^{2})$. \\
\ENDFOR
\end{algorithmic}
\end{algorithm}

\indent
From the processes of Algorithm 11 and Algorithm 12, it can be seen that the $Eagle^{*}$ encryption and $Eagle^{*}$ decryption are mutually inverse, which means that as an encryption algorithm, $Eagle^{*}$ is correct. \\
\indent
In addition, it is easy to see that the calculations are all completed in linear time, so the computational complexity of Algorithm 11 and Algorithm 12 are all $O (KL)$. \\
\indent

\subsubsection{Basic Security Analysis}
\indent
In this chapter, we mainly discuss the security of $Eagle^{*}$ from the following three aspects. \\
\indent
(i) Under the condition of knowing any number of plaintext ciphertext correspondences, cracking the key is equivalent to solving three independent variables $s_{x}, s_{y}, s_{t}$. Through our analysis, to verify the correctness of any two of these three variables, it is necessary to solve a `quintic equation', and we assert that there is no fast solution for this type of equation. Thus, the computational complexity of verifying any two variables is $O(2^{3L-3})$, which is equivalent to the computational complexity that solved the three variables by exhaustive search. \\
\indent
(ii) For attackers who collect a large number of ciphertext plaintext correspondences without cracking the key, and then use statistical or mathematical relationships between plaintext and ciphertext to directly decrypt the new ciphertext, this attack mode often has greater danger. We note that the ciphertexts in the $Eagle^{*}$ encryption algorithm can be viewed as completely independent and random, without any statistical relationship with the plaintext. In addition, theoretically, attackers cannot obtain the algebraic relationship between a single group of plaintext and ciphertext, they can only obtain the nonlinear algebraic expression with ciphertexts as a parameter that satisfies the relationship between adjacent groups of plaintext. Obviously, this directly prevents decrypting new ciphertext from the ciphertext samples. \\
\indent
(iii) The global relationship between the ciphertexts and plaintexts can be seen as constructed through nonlinear operations by the local operation as $s_{x} / \delta x^{2}$. In fact, the local operation $x=s_{x} / \delta x^{2}$ does not satisfy any specific linear relationship between different independent variables $\delta x_{1}, \delta x_{2}$ and different dependent variables $x_{1}, x_{2}$. This means that there is no effecient local analysis technique for $Eagle^{*}$. \\
\indent
To simplify the analysis, in all the following equations, we directly ignore the functions $\phi(x)$ and $\psi(x)$. In fact, these two functions are linear and inverse, and their function is only to make a simple conversion between odd numbers and numbers containing odd numbers of bits with 1. Ignoring them will not have any substantial impact. \\
\indent
In addition, to simplify the unnecessary redundancy in equation expression without affecting the essence of the equation, we can directly ignore the cyclic left shift operation $x^{+(L-1)}$. $\mathbb{E}_(w_{0},w_{1})(m)=m*(w_{0} \oplus w_{1}) \oplus w_{0}^{+(L-1)}$ is represented as $\mathbb{E}_(w_{0},w_{1})(m)=m*(w_{0} \oplus w_{1}) \oplus w_{0}$. \\
\indent
Knowing the plaintexts $P=(P_{1}, P_{2}, ..., P_{K})$ and ciphertexts $C=(C_{1}$, $C_{2}$, ..., $C_{2K+1}$, $\delta x_{1}$, $\delta x_{2}$, ..., $\delta x_{2K+1}$, $\delta y_{1}$, $\delta y_{2}$, ..., $\delta y_{2K+1}$, $z_{1}$, $z_{2}$, ..., $z_{K})$, the problem of cracking the keys $s_{x}, s_{y}, s_{t}$ is equivalent to solving the following system of equations. \\
\begin{equation}
\tag{6.1}
\begin{cases}
((((\delta t_{1} * (x_{1} \oplus y_{1}) \oplus x_{1}) * (u \oplus v) \oplus u) * (x_{2} \oplus y_{2}) \oplus C_{2})   \\
\,\,\,\,\,\,\,\,\,\,\,\,\,\,\,\, * (u \oplus v) \oplus u) * (x_{3} \oplus y_{3}) \oplus (\delta t_{3} * (x_{3} \oplus y_{3}) \oplus x_{3}) \\
\,\,\,\,\,\,\,\,\,\,\,\,\,\,\,\, = C_{3} \\
((((\delta t_{3} * (x_{3} \oplus y_{3}) \oplus x_{3}) * (u \oplus v) \oplus u) * (x_{4} \oplus y_{4}) \oplus C_{4})   \\
\,\,\,\,\,\,\,\,\,\,\,\,\,\,\,\, * (u \oplus v) \oplus u) * (x_{5} \oplus y_{5}) \oplus (\delta t_{5} * (x_{5} \oplus y_{5}) \oplus x_{5}) \\
\,\,\,\,\,\,\,\,\,\,\,\,\,\,\,\, = C_{5} \\
...... \\
\end{cases}
\end{equation}
\indent
Where $s_{t} / \delta t_{2i-1}^{2}=P_{i} \oplus z_{i}$, $x_{i}=s_{x} / \delta x_{i}^{2}$, $y_{i}=s_{y} / \delta y_{i}^{2}$, $u = (s_{x} \oplus s_{y} \oplus 1) / \delta u^{2}$, $v = s_{t} / \delta v^{2}$. \\

\indent
Next, we will demonstrate the security described in (i) by analyzing the following three scenarios. \\
\indent
\textbf{[Case 1]} Verify $s_{x}, s_{t}$. \\
\indent
In order to verify a certain $s_{x}, s_{t}$, each equation in the equation system (6.1) can be regarded as an equation about the unsolvable variable $s_{y}$, After expanding the right multiplication operation in the equations according to the distributive law, we will find that $s_{y}$ in each equation is combined in the form of $* (s_{y} / \delta^2)$ or $(s_{y} / \delta^2)*$, where the most frequent term contains 5 right multiplication operations.\\
\indent
So any equation in the system of equations (6.1) can be regarded as a ``quintic equation" about $s_ {y}$. According to the general conclusion that there is no algebraic solution for quintic equations, we assert that there is no fast solution to the equation about $s_ {y}$ mentioned above, which means that the correctness of each equation can only be verified through exhaustive enumeration of $s_ {y}$. So the computational complexity for verifying $s_ {x}, s_ {t}$ can be considered as $O(2^{L-1})$. \\

\textbf{[Case 2]} Verify $s_{y}, s_{t}$. \\
\indent
Similar to the analysis in Case 1, in order to verify a certain$s_{y}, s_{t}$, any equation in the system of equations (6.1) can be regarded as a ``quintic equation" about $s_ {x} $. We assert that there is no fast solution to the equation about $s_{x}$ mentioned above, which means that the correctness of each equation can only be verified through exhaustive enumeration of $s_{x}$. So the computational complexity for verifying $s_{y}, s_{t}$ can be considered as $O(2^{L-1})$.\\

\textbf{[Case 3]} Verify $s_{x}, s_{y}$. \\
\indent
To verify a certain $s_{x}, s_{y}$, the system of equations (6.1) cannot directly express for $s_{t}$ explicitly. Firstly, we can directly calculate $x_{i}=s_{x} / \delta x_{i}^{2}$ and $y_{i} = s_{y} / \delta y^{2}$ and $u=(s_{x} \oplus s_{y} \oplus 1) / \delta u^{2}$, denote $t_{i}=P_{i} \oplus z_{i}$, the problem of verifying $s_{x}, s_{y}$ can be equivalently transformed into a problem of solving $\delta t_{1}$, thus $v$ can be expressed as $v=t_{1} * \delta t_{1}^{2} / \delta v^{2}$. $\delta t_{3}, \delta t_{5}...$ can be expressed by an implicit equation of the form as $t_{3}=t_{1} * \delta t_{1}^{2} / \delta t_{3}^{2}$, $t_{5}=t_{1} * \delta t_{1}^{2} / \delta t_{5}^{2}$. \\
\indent
Any equation in the system of equations (6.1) can be regarded as a ``quintic equation" about $\delta t_{1}$. We assert that the above equations about $\delta t_{1}$ have no fast solutions, meaning that the correctness of each equation can only be verified by exhaustive enumeration of $\delta t_{1}$. So the computational complexity for verifying $s_{x}, s_{y}$ can be considered as $O(2^{L-1})$.\\

\indent
Based on the discussion of Case 1, Case 2, and Case 3 above, under the condition of knowing the correspondence between plaintexts and ciphertexts, the key is composed of three independent unknown variables. The computational complexity of verifying any two variables is $O(2^{L-1})$, thus the computaional complexity of solving the three variables is $O(2^{3L-3})$, which is equivalent to the computational complexity of completely exhaustively search for $s_{x}, s_{y}, s_{t}$.\\
\\
\indent
Next, we demonstrate the security described in (ii).\\
\indent
Attackers who collect a large number of ciphertext plaintext correspondences to directly decrypt new ciphertext without cracking the key essentially exploit some algebraic relationship between plaintext and ciphertext. After appropriate deduction, we can obtain the relationship between plaintexts and ciphertexts as follows. \\
\begin{equation}
\tag{6.2}
\begin{cases}
((((\delta t_{2i-1} * ((s_{x} / \delta x_{2i-1}^{2}) \oplus (s_{y} / \delta y_{2i-1}^{2})) \oplus (s_{x} / \delta x_{2i-1}^{2})) * (((s_{x} \oplus s_{y} \oplus 1) / \delta u^{2}) \\
\,\,\,\,\,\,\,\,\,\,\,\,\,\,\,\,  \oplus (s_{t} / \delta v^{2})) \oplus ((s_{x} \oplus s_{y} \oplus 1) / \delta u^{2})) * ((s_{x} / \delta x_{2i}^{2}) \oplus (s_{y} / \delta y_{2i}^{2})) \\
\,\,\,\,\,\,\,\,\,\,\,\,\,\,\,\, \oplus C_{2}) * (((s_{x} \oplus s_{y} \oplus 1) / \delta u^{2}) \oplus (s_{t} / \delta v^{2})) \oplus ((s_{x} \oplus s_{y} \oplus 1) / \delta u^{2})) \\     \,\,\,\,\,\,\,\,\,\,\,\,\,\,\,\, *((s_{x} / \delta x_{2i+1}^{2}) \oplus (s_{y} / \delta y_{2i+1}^{2})) \oplus (\delta t_{3} * ((s_{x} / \delta x_{2i+1}^{2}) \oplus (s_{y} / \delta y_{2i+1}^{2}))) \\
\,\,\,\,\,\,\,\,\,\,\,\,\,\,\,\, = C_{3} \\
s_{t} / \delta t_{2i-1}^{2} = P_{i} \oplus z_{i} \\
s_{t} / \delta t_{2i+1}^{2} = P_{i+1} \oplus z_{i+1} \\
\end{cases}
\end{equation}
\indent
It is obvious that through a large number of ciphertext plaintext correspondences, attackers can only obtain algebraic equations satisfied between adjacent groups of plaintext, whose parameters are completely composed of ciphertexts in a nonlinear relationship. That is to say, without cracking the key, attackers can only obtain some non-linear relationship between adjacent groups of plaintext in the most extreme case, but cannot obtain a specific group of plaintext. \\
\\
\indent
Next, we demonstrate the security described in (iii). \\
\indent
Furthermore, we observe that the global relationship between ciphertexts and plaintexts is expressed locally by operation as $s/\delta^{2}$, as shown by  $s_{x} / \delta x_{1}^{2}=x_{1}$, $s_{x} / \delta x_{2}=x_{2}$. For a binary function $F(x,y)$, set $\delta x_{3}=F(\delta x_{1}, \delta x_{2})$, thus $x_{3}=s_{x} / \delta x_{3}^{2}$ can be seen as a binary function about  $x_{1}, x_{2}$ as $x_{3}=G(x_{1}, x_{2})$. Intuitively, $F(x, y)$ and $G(x, y)$ can't be linear functions simultaneously. \\
\indent
Due to the fact that the right multiplication operation defined in this paper does not satisfy the commutative and associative laws, it is obvious that this operation rule can only be regarded as a subset of the operation rules satisfied by traditional multiplication operations. If it can be proven that $F(x, y)$ and $G(x, y)$ cannot both be linear functions for traditional numerical multiplication operations, then for the right multiplication operation described in this paper, it is even more impossible for $F(x, y)$ and $G(x, y)$ to both be linear functions. \\
\indent
Next, we will analyze from the perspective of traditional multiplication operations, assuming$F(x, y)=u_{f}.x+v_{f}.y$ and $G(x,y)=u_{g}.x+v_{g}.y$, where $u_{f} \neq 0$, $v_{f} \neq 0$, $u_{g} \neq 0$, $v_{g} \neq 0$, that is to say, both $F(x, y)$ and $G(x, y)$ are linear functions, after substitution, we obtain $\delta x_{3}=u_{f}.\delta x_{1} + v_{f}. \delta x_{2}$, $x_{3}=u_{g}.x_{1}+v_{g}.x_{2}$, by simple merge we obtain. \\
\begin{equation}
s_{x} / (u_{f}.\delta x_{1}+v_{f}.\delta x_{2})^{2}=u_{g}.(s_{x}/\delta x_{1}^{2})+v_{g}.(s_{x}/\delta x_{2}^{2}) \tag{6.3}
\end{equation}
\indent
By simple algebraic deduction, we obtain \\
\begin{equation}
1/(u_{f} + v_{f}.(\delta x_{1}/\delta x_{2})^{2})=u_{g}+v_{g}.(\delta x_{1}/\delta x_{2})^{2} \tag{6.4}
\end{equation}
\indent
Obviously, $\delta x_{1}/\delta x_{2}$ is a variable, there is no fixed parameters $(u_{f}$, $v_{f}$, $u_{g}$, $v_{g})$ that make (6.4) hold for any $\delta x_{1}/\delta x_{2}$. That is to say, from the perspective of traditional multiplication operations, there is no binary function $F(x, y), G(x, y)$ that both satisfies a linear relationship, therefore, from the perspective of the right multiplication operation defined in this paper, there is no binary function $F(x, y), G(x, y)$ that both satisfies a linear relationship too. \\
\indent
The absence of the linear binary functions $F(x, y)$ and $G(x, y)$ mentioned above also indicates that without cracking the key, attackers cannot derive the linear relationship of local variables $x_{i}, x_{j}$ from any linear relationship between two ciphertexts $\delta x_{i}, \delta x_{j}$. Similarly, attackers are unable to derive the linear relationship of local variables $y_{i}, y_{j}$ from any linear relationship between two ciphertexts $\delta y_{i}, \delta y_{j}$. Due to the fact that $C_{i}$ can be mapped to $\delta t_{i}$, attackers are also unable to derive the linear relationships of local variables $t_{i}, t_{j}$ ($t_{i}=s_{t} / \delta t_{i}^{2}$, $t_{j}=s_{t} / \delta t_{j}^{2}$) from any linear relationship between any two ciphertexts $C {i}, C {j}$.\\

\subsubsection{Linear attack analysis to $Eagle^{*}$ encryption algorithm}
\indent
Linear attack is a very effective attack method proposed by M. Matsui\cite{REF2} at the European Cryptology Conference in 1993. Later, scholars quickly  discovered  that  the   linear   attacks  are  applicable  to  almost  all   block encryption algorithms, and linear attacks have became the main attacks for block encryption algorithms. Various new attacks based on linear attacks are constantly being proposed.\\
\indent
The  core  idea  of  linear  attack  is  to  take  the  nonlinear  transformation  in  the cryptographic algorithm, such as the linear approximation of the S-box, and then extend the linear approximation to the linear approximation of the round function, and then connect these linear approximations to obtain a   linear    approximation of the entire cryptographic algorithm, finally a    large   number   of    known plaintext-ciphertext  pairs encrypted with the same  key are used to exhaustively obtain the plaintext and even the key.\\
\indent
We have noticed that the reason why linear attacks have become an effective attack for block encryption algorithms is that when the key is known, there is a certain implicit linear relationship between the ciphertext and the plaintext. By analyzing the   known  plaintext-ciphertext   pairs,  some  effective  linear   relations  can  be obtained, and some bits of the key can be guessed.\\
\indent
In the $Eagle^{*}$ encryption algorithm, the entire encryption system is divided into three independent parts, each with its own independent key and parameters. The ciphertext of the encryption system that includes the original plaintext is generated by mixing the original plaintext with random parameters, without knowing the random parameters, ciphertext can be regarded as completely independent of the original plaintext, which means that with random parameter changes, ciphertext can equally probability traverse any text in the ciphertext space. And the two other parts of the ciphertext is completely generated by random parameters, completely unrelated to the original plaintext. This can be expressed as \\
\begin{equation}
\notag
\begin{aligned}
Pr(C_{i}=c_{i} | P_{i}=p_{i})=\frac {1} {2^{L-1}}   \\
Pr(\delta x_{i}=sx_{i} | P_{i}=p_{i})=\frac {1} {2^{L-1}}  \\
Pr(\delta y_{i}=sy_{i} | P_{i}=p_{i})=\frac {1} {2^{L-1}}  \\
Pr(z_{i}=sz_{i}| P_{i}=p_{i})=\frac {1} {2^{L}}  \\
Pr(\delta u=du | P_{i}=p_{i})=\frac {1} {2^{L-1}}  \\
Pr(\delta v=dv | P_{i}=p_{i})=\frac {1} {2^{L-1}}  \\
Pr(\delta t_{2K+1}=dt_{2K+1} | P_{i}=p_{i})=\frac {1} {2^{L-1}}  \\
\end{aligned}
\end{equation}

\indent
\subsubsection{Differential attack analysis to $Eagle^{*}$ encryption algorithm}
Differential attack was proposed by Biham and Shamir\cite{REF6} in 1993, it is a chosen- plaintext attack.  Its core  idea  is to obtain  key  information  by analyzing specific plaintext and ciphertext differences.\\
\indent
The essence of a differential attack is to track the ``difference'' of the plaintext pair, where the ``difference'' is defined by the attacker according to the target, which can be an exclusive XOR operation or other target values. \\
\indent
For  the   $Eagle^{*}$   encryption  algorithm,  suppose  the differential attacker chooses two specific plaintexts  $P_{1}$ and $P_{2}$, their difference is  $\Delta$, that is  $P_{2}  = P_{1}  + \Delta$, Since the ciphertexts in the second part and the third part are completely unrelated to the plaintext, only the ciphertext in the first part will be analyzed here, the corresponding ciphertexts are   $C_{1}$     and  $C_{2}$ , and the difference  between the ciphertexts  is  $\varepsilon$ as $C_{2} = C_{1} + \varepsilon$. Since $C_{1}$    and   $C_{2}$    are completely random, it is uncertain whether the difference  $\varepsilon$  of  the ciphertext  is caused  by randomness of $(x_{1}, y_{1}, x_{2}, y_{2}, u)$  or  the uncertainty of  the   plaintext $(P_{1}, P_{2})$.  Furthermore, for any key $(\delta_{1}, \delta_{2})$,  $x_{1} * \delta_{1} * \delta_{1}$,  $y_{1} * \delta_{2} * \delta_{2}$ subject to the uniform distribution, which can be denoted as \\
\begin{equation}
\notag
\begin{aligned}
Pr(C_{1}  = c_{1}, ... ,C_{2K+1} = c_{2K+1} |\\
 (P_{1}  = p_{1},..., P_{K}=p_{K}, W=(s_{x}, s_{y}, s_{t})) \\
	= 1/ 2^{(2K+1)(L-1)} \\
\end{aligned}
\end{equation}
\begin{equation}
\notag
\begin{aligned}
Pr(\delta x_{1}  = dx_{1}, ... ,\delta x_{2K+1} = dx_{2K+1}) |\\
 (P_{1}  = p_{1},..., P_{K}=p_{K}, W=(s_{x}, s_{y}, s_{t})) \\
	= 1/ 2^{(2K+1)(L-1)} \\
\end{aligned}
\end{equation}
\begin{equation}
\notag
\begin{aligned}
Pr(\delta y_{1}  = dy_{1}, ... ,\delta y_{2K+1} = dy_{2K+1} |\\
 (P_{1}  = p_{1},..., P_{K}=p_{K}, W=(s_{x}, s_{y}, s_{t})) \\
	= 1/ 2^{(2K+1)(L-1)} \\
\end{aligned}
\end{equation}
\begin{equation}
\notag
\begin{aligned}
Pr(z_{1}  = sz_{1}, ... ,z_{K} = sz_{K} |\\
 (P_{1}  = p_{1},..., P_{K}=p_{K}, W=(s_{x}, s_{y}, s_{t})) \\
	= 1/ 2^{K(L-1)} \\
\end{aligned}
\end{equation}
\begin{equation}
\notag
\begin{aligned}
Pr(\delta u = du, \delta v = dv, \delta t_{2K+1}=dt_{2K+1} |\\
 (P_{1}  = p_{1},..., P_{K}=p_{K}, W=(s_{x}, s_{y}, s_{t})) \\
	= 1/ 2^{3(L-1)} \\
\end{aligned}
\end{equation}
\indent
That is to say, for any specific plaintext  $P_{1}$     and  $P_{2}$     selected by the attacker, after being encrypted with the same key, the corresponding block ciphertexts  $(C_{1}$, $C_{2}$, ..., $C_{2K+1})$, $(\delta x_{1}$, $\delta x_{2}$, ..., $\delta x_{2K+1})$, $(\delta y_{1}$, $\delta y_{2}$, ..., $\delta y_{2K+1})$, $(z_{1}$, $z_{2}$, ..., $z_{K})$, $\delta u, \delta v, \delta t_{2K+1}$  are completely random, and every possible value in the ciphertext space appears with  equal   probability.  The  attacker   has   no  way  to  capture  the   propagation characteristics of the ``difference'' in the plaintext.\\
\indent

\begin{appendices}
\appendix
\section{General quintic equation with Eagle right multiplication}
\indent
As introduced in chapter 4.1, We have constructed a right multiplication operation, denoted as $m * \delta$. and also constructed a corresponding right division operation, denoted as $s / \delta$. \\
\indent
Similar to the traditional definition of power operations, we can also define right power operations as follows \\
\begin{equation}
(m_{1} \circ m_{2} \circ ... \circ m_{n}) * \delta^{n}=m_{1} * \delta * m_{2} * \delta * ... * m_{n} * \delta \tag{A.1}
\end{equation}
\indent
(A.1) represents performing $n$ consecutive right multiplication operations on the parameter $\delta$ using $(m_{1}, m_{2}, ..., m_{n})$. \\
\indent
Similarly, right root operations can also be defined.\\
\begin{equation}
(m_{1} \circ m_{2} \circ ... \circ m_{n}) * \delta^{n}=s \tag{A.2}
\end{equation}
\indent
The root of equation (A.2) can be expressed as \\
\begin{equation}
\delta=\sqrt[n]{s[m_{1} \circ m_{2} \circ ... \circ m_{n}]} \tag{A.3}
\end{equation}
\indent
So, the general quintic equation for Eagle right multiplication operation can be expressed as. \\
\begin{equation}
\tag{A.4}
\begin{aligned}
(a_{1} \circ a_{2} \circ a_{3} \circ a_{4} \circ a_{5}) * \delta^{5} \oplus  (b_{1} \circ b_{2} \circ b_{3} \circ a_{4}) * \delta^{4} \\
\oplus (c_{1} \circ c_{2} \circ c_{3})*\delta^{3} \oplus (d_{1} \circ d_{2})*\delta^{2} \oplus e*\delta = f
\end{aligned}
\end{equation}
\indent
Similar to the conclusion that traditional general quintic equations have no algebraic solutions, the quintic equation defined as (A.4) also has a similar conclusion, as follows \\
\indent
\textbf{[Theorem A.1]} The equation defined in (A.4) has no algebraic solution. \\
\begin{proof}
\indent
Proof by contradiction, we suppose that there exists an algebraic solution for (A.4) expressed as $h$, such that $\delta=h(a_{1}, ..., a_{5}, b_{1}, ..., b_{4}, c_{1}, ..., c_{3}, d_{1}, d_{2}, e, f)$. Where $f$ is composed of parameters $(a_{1}, ..., a_{5}, b_{1}, ..., b_{4}, c_{1}, ..., c_{3}, d_{1}, d_{2}, e, f)$ that undergo a finite number of $\oplus$ operations, right multiplication operations, right division operations, right power operations, right root operations.  \\
\indent
Now consider another traditional quintic equation as follows. \\
\begin{equation}
\tag{A.5}
\begin{aligned}
\bar{a_{1}}.\bar{a_{2}}.\bar{a_{3}}.\bar{a_{4}}.\bar{a_{5}}.\bar{\delta}^{5}+\bar{b_{1}}.\bar{b_{2}}.\bar{b_{3}}.\bar{b_{4}}.\bar{\delta}^{4} \\
+\bar{c_{1}}.\bar{c_{2}}.\bar{c_{3}}.\bar{\delta}^{3}+\bar{d_{1}}.\bar{d_{2}}.\bar{\delta}^{2}+\bar{e}.\bar{\delta}=\bar{f}
\end{aligned}
\end{equation}
\indent
We perform the following regular transformations on all symbols in the expression $h(a_{1}, ..., a_{5}, b_{1}, ..., b_{4}, c_{1}, ..., c_{3}, d_{1}, d_{2}, e, f)$. \\
\indent
(i) Change the symbol $a_{i}$ to $\bar{a_{i}}$, change the symbol $b_{i}$ to $\bar{b_{i}}$, change the symbol $c_{i}$ to $\bar{c_{i}}$, change the symbol $d_{i}$ to $\bar{d_{i}}$, change the symbol $e$ to $\bar{e}$, change the symbol $f$ to $\bar{f}$. \\
\indent
(ii) Change the symbol $\oplus$ to $\pm$. \\
\indent
(iii) Change the symbol $a * b$ to $a.b$. \\
\indent
(iv) Change the symbol $a / b$ to $a/b$. \\
\indent
(v) Change the symbol $(m_{1} \circ m_{2} \circ ... \circ m_{n}) * \delta^{n}$ to $m_{1}.m_{2}.....m_{n}.\delta^{n}$. \\
\indent
(vi) Change the symbol $\sqrt[n]{s[m_{1} \circ m_{2} \circ ... \circ m_{n}]}$ to $\sqrt[n]{s/(m_{1}.m_{2}.....m_{n})}$. \\
\indent
(vii) Change the constant to a universal constant symbol (after constant replacement, although the specific constant value after replacement is currently unknown, such a constant must be a certain determined value after replacement). \\
\indent
The expression after transformation is $\bar{h}(\bar{a_{1}}$, ..., $\bar{a_{5}}$, $\bar{b_{1}}$, ..., $\bar{b_{4}}$, $\bar{c_{1}}$, ..., $\bar{c_{3}}$, $\bar{d_{1}}$, $\bar{d_{2}}$, $\bar{e}$, $\bar{f})$. \\
\indent
Since replacing $\delta$ in (A.4) with $h(a_{1}, ..., a_{5}, b_{1}, ..., b_{4}, c_{1}, ..., c_{3}, d_{1}, d_{2}, e, f)$ makes the equation hold, it can be seen from the homogeneity of rules (i)-(vii) that replacing $\bar{\delta}$ in (A.5) with $\bar{h}(\bar{a_{1}}, ..., \bar{a_{5}}, \bar{b_{1}}, ..., \bar{b_{4}}, \bar{c_{1}}, ..., \bar{c_{3}}, \bar{d_{1}}, \bar{d_{2}}, \bar{e}, \bar{f})$ also makes the equation hold. \\
\indent
It should be pointed out that the $\pm$ symbol in $\bar{h}$ does not mean that replacing $\pm$ with $+$ or $-$ will make the equation hold, but rather that each $\pm$ can only be replaced with a specific $+$ or $-$ to make the equation hold, such replacement is definitely feasible. \\
\indent
According to the conclusion that there are no algebraic solutions for general quintic equations, it can be concluded that such $\bar{h}$ does not exist, and therefore $h$ satisfying (A.4) also does not exist. \\
\end{proof}
\indent
Theorem A.1 states that although the general quintic equation defined by Eagle right multiplication operation has no algebraic solution, it cannot be ruled out that it has some approximate numerical expression, which requires the following theorem. \\
\indent
\textbf{[Theorem A.2]} The right multiplication $m * \delta$ can't establish an ``order" relationship $\prec$, such that for $\delta_{1} \prec \delta_{2}$, the relationship $m * \delta_{1} \prec m * \delta_{2}$ always holds for any $m$. \\
\begin{proof}
\indent
Proof by contradiction, we assume that there exists an order relationship $\prec$, such that for $\delta_{1} \prec \delta_{2}$, the relationship $m * \delta_{1} \prec m * \delta_{2}$ holds for any $m$. \\
\indent
After sorting all $\delta$ in the order of $\prec$, the sequence is denoted as $\delta_{1} \prec \delta_{2} \prec ... \prec \delta_{2^{L-1}}$. \\
\indent
For any $m$, $m * \delta_{1} \prec m * \delta_{2} \prec ... \prec m * \delta_{2^{L-1}}$ holds. \\
\indent
Since the sequence $\delta_{1} \prec \delta_{2} \prec ... \prec \delta_{2^{L-1}}$ and the sequence $m * \delta_{1} \prec m * \delta_{2} \prec ... \prec m * \delta_{2^{L-1}}$ are the same sequence, it can be deduced that $m * \delta_{i}=\delta_{i}$, which is obviously incorrect. \\
\end{proof}
\indent
Due to the inability to establish an effective order relationship based on Eagle right multiplication operation, it is impossible to establish some numerical correspondence, and thus it's unable to measure the approximation of a solution to the equation (A.4). So we can conclude that there is no approximate solution for equations of type (A.4). \\

\section{More about brute-force searching}
\indent
In the computational complexity theory, a problem is deemed difficult if it cannot be solved within polynomial time. Furthermore, if a certain internal variable with a problem can only be solved through brute-force searching, we can always construct the domain size of this variable to be exponential in scale, resulting in the problem having exponential complexity. \\
\indent
Can the internal variable of a problem only be solved through brute-force searching? We summarize this problem into the following three levels semantically: \\
\indent
(i) When a variable cannot be expressed with symbols in finite length for any constraint condition, its value cannot be solved using effective general algorithms other than brute-force searching. \\
\indent
(ii) When a variable needs to be expressed in any constraint condition which the length of the symbols used must exceed the size of its domain, the computational complexity for solving this variable should not be lower than that of brute-force searching.  \\
\indent
(iii) When there are inherent obstacles in the constraints of a variable itself, there is no better way to solve this variable than brute-force searching. \\
\indent
In fact, in the Eagle encryption algorithm constructed in this paper, the problem of cracking the key is equivalently transformed into a general quintic equation with one variable as the unknown variable and the other variables as coefficients. According to the conclusion that general quintic equations have no algebraic solutions, and the right multiplication algorithm constructed in this paper does not have a numerical algorithm for solving approximate values, thus this proves that any constraint condition of the variable in the equation cannot be expressed by finite length symbols of other variables as coefficients. This can also be equivalently understood as the inability to establish equations that satisfy the constraint conditions between variables as coefficients through finite length expressions, so naturally there is no universal algorithm other than brute-force searching to solve these variables. \\

\end{appendices}

\bibliographystyle{plain}

\end{CJK}

\end{document}